\documentstyle[twoside,epsf]{article}
\newcount\Mac  \Mac=0  
\newcommand{\ifMac}[2]{\ifnum\Mac=1 #1 \else #2 \fi}
\def\putps(#1,#2)(#3,#4)#5#6{\ifnum\Mac=1 \put(#1,#2){\special{picture #5}}
\else  \put(#3,#4){\includegraphics{#6}} \fi}
\def\putps(#1,#2)(#3,#4)#5#6{\put(#1,#2){\includegraphics{#6}}
\put(#3,#4){\includegraphics{#6}} }
\def\Red  {}
\def\Black{}
\def\Blue {}

\def\wtil{\widetilde}

\newcommand{\GeV}{\,{\rm GeV}}
\newcommand{\eV}{\,{\rm eV}}

\newcommand{\NP}{Nucl. Phys.}

\newcommand{\PL}{Phys. Lett.}
\newcommand{\PR}{Phys. Rev.}

\newcommand{\hu}{{H_{\rm u}}}
\newcommand{\hd}{{H_{\rm d}}}
\newcommand{\mhu}{m_{H_{\rm u}}}
\newcommand{\mhd}{m_{H_{\rm d}}}

\newcommand{\eq}[1]{~(\ref{eq:#1})}

\newcommand{\fig}[1]{~\ref{fig:#1}}
\newcommand{\MGUT}{M_{\rm GUT}}

\def\Ord{{\cal O}}  
 \def\Tr{\mathop{\rm Tr}}
\def\circa#1{\,\raise.3ex\hbox{$#1$\kern-.75em\lower1ex\hbox{$\sim$}}\,}
\makeatletter
\def\art{\@ifnextchar[{\eart}{\oart}}
\def\eart[#1]#2#3#4#5#6{{\rm #2}, {\em #3 \rm #4} {\rm (#6) #5 ({\rm #1})}}
\def\hepart[#1]#2{{\rm #2, #1}}
\newcommand{\oart}[5]{{\rm #1}, {\em #2 \rm #3} {\rm (#5) #4}}

\newcounter{alphaequation}[equation]
\def\thealphaequation{\theequation\hbox to
0.6em{\hfil\alph{alphaequation}\hfil}}
\def\eqnsystem#1{
\def\@eqnnum{{\rm (\thealphaequation)}}
\def\@@eqncr{\let\@tempa\relax \ifcase\@eqcnt \def\@tempa{& & &} \or
  \def\@tempa{& &}\or \def\@tempa{&}\fi\@tempa
  \if@eqnsw\@eqnnum\refstepcounter{alphaequation}\fi
\global\@eqnswtrue\global\@eqcnt=0\cr}
\refstepcounter{equation} \let\@currentlabel\theequation \def\@tempb{#1}
\ifx\@tempb\empty\else\label{#1}\fi
\refstepcounter{alphaequation}
\let\@currentlabel\thealphaequation
\global\@eqnswtrue\global\@eqcnt=0 \tabskip\@centering\let\\=\@eqncr
$$\halign to \displaywidth\bgroup \@eqnsel\hskip\@centering
$\displaystyle\tabskip\z@{##}$&\global\@eqcnt\@ne
\hskip2\arraycolsep\hfil${##}$\hfil& \global\@eqcnt\tw@\hskip2\arraycolsep
$\displaystyle\tabskip\z@{##}$\hfil
\tabskip\@centering&\llap{##}\tabskip\z@\cr}
\def\endeqnsystem{\@@eqncr\egroup$$\global\@ignoretrue} \makeatother

\newcommand{\riga}[1]{\noalign{\hbox{\parbox{\columnwidth}{#1}}}\nonumber}

\def\L{{\cal L}}
\def\O{{\cal O}}

\def\Z{{\cal Z}}
\def\li2{{\rm Li}_2}

\def\roughly#1{\,\,\raise.3ex\hbox{$#1$\kern-.75em\lower1ex\hbox{$\sim$}}\,\,}

\def\bea{\begin{eqnarray}}
\def\eea{\end{eqnarray}}
\def\beq{\begin{equation}}
\def\eeq{\end{equation}}
\def\eq{\end{equation}}
\def\to{\rightarrow}

\def\fo{\hbox{{1}\kern-.25em\hbox{l}}}

\def\beq{\begin{equation}}
\def\eeq{\end{equation}}
\def\eq{\end{equation}}
\def\to{\rightarrow}

\newcommand{\newc}{\newcommand}

\newc{\lcal}{\int {\cal L}dt}
 
\newc{\LSP}{{\chi^0_1}}
\newc{\stauR}{{\tilde \tau_R}}
\newc{\stau}{{\tilde \tau_1}}
\newc{\mstop}{m_{\tilde{t}}}
\newc{\mHpm}{m_{H^\pm}}
\newc{\gsim}{\lower.7ex\hbox{$\;\stackrel{\textstyle>}{\sim}\;$}}
\newc{\lsim}{\lower.7ex\hbox{$\;\stackrel{\textstyle<}{\sim}\;$}}
\newc{\ie}{{\it i.e.}}          
\newc{\etal}{{\it et al.}}
\newc{\eg}{{\it e.g.}}          
\newc{\kev}{\hbox{\rm\,keV}}            
\newc{\mev}{\hbox{\rm\,MeV}}            
\newc{\gev}{\hbox{\rm\,GeV}}            
\newc{\tev}{\hbox{\rm\,TeV}}
\newc{\xpb}{\hbox{\rm\, pb}}
\newc{\xfb}{\hbox{\rm\, fb}}

\def\beq{\begin{equation}}
\def\eeq{\end{equation}}
\def\bea{\begin{eqnarray}}
\def\eea{\end{eqnarray}}
\def\slashchar#1{\setbox0=\hbox{$#1$}           
   \dimen0=\wd0                                 
   \setbox1=\hbox{/} \dimen1=\wd1               
   \ifdim\dimen0>\dimen1                        
      \rlap{\hbox to \dimen0{\hfil/\hfil}}      
      #1                                        
   \else                                        
      \rlap{\hbox to \dimen1{\hfil$#1$\hfil}}   
      /                                         
   \fi}                                         %
\catcode`@=11
\long\def\@caption#1[#2]#3{\par\addcontentsline{\csname
  ext@#1\endcsname}{#1}{\protect\numberline{\csname
  the#1\endcsname}{\ignorespaces #2}}\begingroup
    \small
    \@parboxrestore
    \@makecaption{\csname fnum@#1\endcsname}{\ignorespaces #3}\par
  \endgroup}
\catcode`@=12

\def\jfig#1#2#3{
 \begin{figure*}[t]
 \centering
 \epsfysize=3.0in
 \epsffile{#2}
 \caption{\em #3}
 \label{#1}
 \end{figure*}}

\begin{document}
\centerline{\bf Dec.\ 1999 \hfill    IFUP--TH/54--99}
\centerline{\bf hep-ph/9912390 \hfill SNS--PH/99--16} 
\centerline{\bf \hfill UCD-99-25} \vspace{1cm}
\centerline{\LARGE\bf\Red Phenomenology of deflected anomaly-mediation}\vspace{0.2cm}
\bigskip\bigskip\Black
\centerline{\large\bf Riccardo Rattazzi} \vspace{0.3cm}
\centerline{\em INFN and Scuola Normale Superiore, I-56100 Pisa, Italia}\vspace{0.3cm}
\centerline{\large\bf Alessandro Strumia}\vspace{0.3cm}
\centerline{\em Dipartimento di fisica, Universit\`a di Pisa and INFN,  I-56126 Pisa, Italia}\vspace{0.3cm}
\centerline{\large  and }\vspace{0.3cm}
\centerline{\large\bf James D. Wells}\vspace{0.3cm}
\centerline{\em Physics Department, University of California, Davis, CA 95616}

\bigskip\bigskip\Blue

\centerline{\large\bf Abstract}
\begin{quote}\large\indent
We explore the phenomenology of a class of
models with anomaly-mediated supersymmetry breaking. These models
retain the successful flavor 
properties of the minimal scenario while avoiding the tachyons.
The mass spectrum is predicted in terms of a few parameters.
However various qualitatively different
spectra are possible, often strongly different from the
ones usually employed to explore capabilities of new accelerators. 
One stable feature is the limited spread of the spectrum, so that
squarks and gluinos could be conceivably  produced at TEVII.
The lightest superpartner of standard particles is often a charged slepton 
or a neutral higgsino. It behaves as a stable particle in collider
experiments but it decays at or before nucleosynthesis.
We identify the experimental signatures at hadron colliders that can help 
distinguish this scenario from the usual ones. 

\end{quote}
\Black\vspace{1cm}


\section{Introduction}

The origin of supersymmetry breaking is the central issue in the 
construction of a realistic supersymmetric extension of
the Standard Model (SM). If supersymmetry is to be of any 
relevance to the 
hierarchy problem the sparticle masses should be smaller than about a TeV.
Then, flavor violating processes mediated by virtual sparticles constrain
their masses to preserve flavor to a high degree. One main
goal of model building is to provide flavor symmetric soft terms in
a simple and natural way. Gauge mediated supersymmetry breaking (GMSB) 
\cite{gmsb} represents
an elegant solution to this problem: soft terms are calculable
and are dominated by a flavor symmetric contribution due to gauge interactions.
Supergravity, on the other hand, provides perhaps the simplest way
to mediate supersymmetry breaking~\cite{gravmed}. However, in the absence 
of a more  fundamental theory, soft terms are not calculable
in supergravity, so there is little control on their flavor structure.
More technically, one could say that soft terms are dominated by ``extreme 
ultraviolet'' dynamics in supergravity and consequently are sensitive
to all possible new sources of flavor violation, not just the
``low-energy'' Yukawa couplings. This can be  considered a 
generic problem 
of soft terms mediated by supergravity. Various solutions have
been suggested, including special string inspired scenarios (dilaton
dominance) and horizontal symmetries. 

Recently, important progress
has been made in our understanding of a class of calculable
quantum effects in supergravity
\cite{RS,GLMR}. These effects can be characterized
as the {\it pure} supergravity contribution to soft terms. This is because
they are simply determined by the vacuum expectation value of the auxilliary
scalar field $F_\phi$ in the graviton supermultiplet. The couplings of 
$F_\phi$ to 
the minimal supersymmetric standard model (MSSM)
are a purely quantum effect dictated by the conformal anomaly. The resulting 
{\it anomaly mediated}  contribution to sparticle masses is of order 
$\alpha F_\phi/4\pi\sim \alpha m_{3/2}/4\pi$. In a generic supergravity 
scenario, this calculable effect would only represent
a negligible correction to the uncalculable $\sim m_{3/2}$ tree level terms.
However it is consistent to consider a situation where Anomaly Mediation
(AM) is the leading
effect. Indeed, as pointed out by Randall and Sundrum~\cite{RS}, this
may happen in an  extra-dimensional scenario, for example,
when the MSSM lives on
a 3-brane, while the hidden sector lives on a brane that is far-away in a bulk
where {\it only} gravity propagates. Recently an explicit realization of 
this setup has been given in ref.~\cite{LS}.
A more conventional situation, where the anomaly mediated contribution to just
the gaugino masses and $A$-terms dominates, is dynamical hidden sector models
without singlets~\cite{GLMR}. 
Various technical aspects of AM have been further discussed
in Refs.~\cite{PR,CLMP,BMP}, the latter of which gives a more formal
derivation 
along with a comparison to previous computations of quantum contributions
to soft terms~\cite{KL}.

In  pure Anomaly Mediation
sfermion masses are dominated by an infrared contribution, 
so they are only sensitive to the sources of flavor violation
that are relevant at low energy, as encoded in the fermion masses and
CKM angles of the SM. Therefore AM, like the SM,  satisfies natural 
flavor conservation. Sfermion masses are in practice 
family independent, since the gauge contributions dominate, like in GMSB.
Unfortunately, this is not the full story: 
flavor is fine but the squared
slepton masses are predicted to be negative.

Various attempts have been made to save the situation.
In principle adding an extra supergravity contribution ruins 
predictivity. Nevertheless, if one assumes that some
unspecified flavor universal contribution lifts the sleptons,
then the low-energy phenomenology is quite 
peculiar~\cite{GGW,fengetal}.
Other proposals involve extra fields at, or just above, the weak 
scale~\cite{KSS,CLMP}.
In this paper we will focus on the idea of ref.~\cite{PR},
which we outline below.

The fact that AM provides a special Renormalization Group (RG) trajectory
where all unwanted ultraviolet (UV) effects on soft terms decouple is very 
suggestive. 
Indeed, in order to solve the supersymmetric flavor problem,
it would be enough to remain on this trajectory only down to a scale
$M_0$ somewhat below the scale of flavor. 
In ref.~\cite{PR} it was pointed out that a theory can be kicked
off the AM trajectory when an intermediate theshold is governed
by the vacuum expectation value (VEV)
 of a field $X$ that is massless in the supersymmetric limit.
This does not truly violate the UV insensitivity of AM, since the low
energy theory is not just the MSSM but contains also 
the modulus $X$. While this field is coupled to the MSSM only by $1/X$
suppressed operators, its presence affects the  soft masses in a relevant way. 
 Ref.~\cite{PR} used this remark to build a realistic class
of models, with flavor universal and positive sfermion masses. The
intermediate threshold  is given by a messenger sector similar to that
of GMSB models. However the sparticle spectrum of these models
strongly differs from both GMSB and conventional supergravity. 
Indeed the prediction for gaugino mass ratios is also distinguished
from ``minimal'' AM. 
The most important features of the spectrum are a reduced hierarchy between 
coloured
sparticles and the rest, and the lightest spartner being either
a slepton or a higgsino-like neutralino. The 
lightest supersymmetric particle (LSP) is the fermionic 
partner $\chi$ of the modulus $X$, so the lightest sparticle in the MSSM 
can be charged.

The purpose of the present paper is to study the implications
of these novel features in collider physics and cosmology.
It is organized as follows. In section 2 we recall the building blocks of 
the model and the corresponding high-scale boundary conditions for soft terms.
In section 3 we study the low-energy 
spectrum and consider the constraints from 
electroweak symmetry breaking. In section 4 we focus on 
the signatures at both TEVII and LHC and draw a comparison to those
of GMSB and minimal supergravity (mSUGRA).
Supersymmetric corrections to rare processes are studied in section~5.
In section 6 we discuss the NLSP decays and the bounds
on it placed by big-bang nucleosynthesis.
Section 7 contains our  conclusions.
In appendix A we write the one-loop RG evolution for the soft terms 
in terms of a minimal number of `semi-analytic' functions,
starting from the most general boundary conditions.



\section{The model}

Anomaly Mediated soft terms can be defined in a very simple operational 
way. Consider first any model in the supersymmetric
limit and assign $R$-charge $2/3$ to all its chiral matter 
superfields. Notice that in general this is not a true symmetry.
For instance, in the superpotential only the trilinear couplings 
are invariant.  Consider then the introduction of
a spurion (classical external field) $\phi$ with $R$-charge 2/3
and scaling-dimension 1, and couple it to the original lagrangian in order
to make it formally both $R$ and scale invariant. For instance for a generic 
superpotential $W(Q)$ we have
\beq
W(Q)=M_1Q^2+\lambda Q^3 +{1\over M_{-1}}Q^4+\dots\,\to\,
  M_1\phi Q^2+\lambda Q^3 +{1\over M_{-1}\phi}Q^4+\dots =\phi^3W(Q/\phi).
\label{defw}
\eeq
When the choice $\phi=1+\theta^2 F_\phi$ is made, some special soft terms
are generated: they are proportional to the dimension of the original
superpotential coupling. Notice that they vanish for a purely
cubic $W$. The same game can be played with the gauge interaction terms.
Like for Yukawas, the coupling to $\phi$  is absent because gauge interactions 
are scale invariant and $R$ symmetric at tree level.
So in a theory with only gauge and Yukawa couplings no soft term arises at
tree level. However, a  coupling to $\phi$ arises at the quantum level due
to anomalous breaking of  scale (and $R$) invariance. 
Indeed, in order to formally restore the two symmetries
one should also couple the regulator Lagrangian to $\phi$. For instance
in supersymmetric QED the Pauli-Villars mass should be multiplied by a factor $\phi$,
like in eq.~(\ref{defw}). The quantum dependence on $\phi$ can be effectively
accounted for by considering superfield matter wave functions and gauge 
couplings
\cite{aglr}
\beq
\Z_i(\mu)=Z\left({\mu \over {\sqrt{\phi\phi^\dagger}}}\right )\quad
\quad R(\mu)=g^{-2}\left({\mu \over {\sqrt{\phi\phi^\dagger}}}\right )
\label{superfield}
\eeq
where $Z_i(\mu) $ and $g^2(\mu )$ are the running 
parameters in the supersymmetric limit. Eq. \ref{superfield}
is derived by noticing that the quantity $\mu/
{\sqrt{\phi\phi^\dagger}}$ is the only scale and $R$ invariant
combination of $\mu$ and $\phi$~\cite{RS,GLMR}. By eq.~(\ref{superfield})
the $A$-terms, scalar and gaugino masses are
\begin{eqnsystem}{sys:gen}
&\parbox{7cm}{$A_{ijk}(\mu)=-{1\over 2}\left( \gamma_i(\mu)+\gamma_j(\mu)
  +\gamma_k(\mu)\right) F_\phi$}
& \gamma_i=\frac{d\ln Z_i}{d\ln \mu}  \\
&\parbox{7cm}{$m_i^2(\mu)=-{1\over 4 }\dot\gamma_i(\mu) |F_\phi|^2$} &
\dot\gamma_i=\frac{d\gamma_i}{d\ln \mu}\\
&\parbox{7cm}{$\displaystyle m_\lambda(\mu)=\frac{\beta(g^2(\mu))}{2g^2(\mu)}F_\phi$} &
\beta=\frac{d g^2}{d\ln\mu}
\label{am}
\end{eqnsystem}
where $A_{ijk}$ is the dimensionful scalar-Yukawa analogous 
to the Yukawa coupling $\lambda_{ijk}$. The pure 
gauge contribution to scalar masses is proportional to $-\beta(g^2)$, which is 
positive for asymptotically free gauge theories and negative otherwise.
In the MSSM neither $SU(2)_L$ nor $U(1)_Y$ is asymptotically free.
So the slepton squared masses, which are dominated  by the $SU(2)_L\times 
U(1)_Y$  contribution, are negative and the model is ruled out.

The models constructed in ref.~\cite{PR} eliminate the
tachyons while preserving the successful flavor properties of AM. 
In these models $n$ flavors 
of `messengers' $\Psi_i$, $\bar \Psi_i$ in the ${\bf 5 +\bar 5}$ of $SU(5)$
and a singlet $X$ are added to the MSSM fields.
These fields interact via the superpotential
\beq
W_{\rm mess}=\lambda_\Psi X\Psi_i\bar \Psi_i
\label{xpsipsi}
\eeq
so the basic structure is that of GMSB models. However it is assumed
that soft terms are generated by AM already in supergravity.
We are interested in a situation where $X$ gets a large VEV so that
the messengers are ultra-heavy. If $\langle X\rangle$ were fixed by 
supersymmetric dynamics, for example by a superpotential $W(X)$, then
the relation $F_X/\langle X\rangle=F_\phi$ would hold in the presence
of the spurion $\phi$. The messenger supermultiplets would then be split,
and upon integrating them out a gauge-mediated correction
to the sparticle masses would arise. By the relation 
$F_X/\langle X\rangle=F_\phi$, this correction would precisely adjust
the soft terms to the AM trajectory of the low-energy theory,
 {\it i.e} to the beta functions of the theory without messengers. This
is just an example of the ``celebrated'' decoupling of heavy
thresholds in AM. 

\begin{figure*}[t]
\begin{center}
\begin{picture}(17.7,6)
\putps(0,0)(0,0){fRGE}{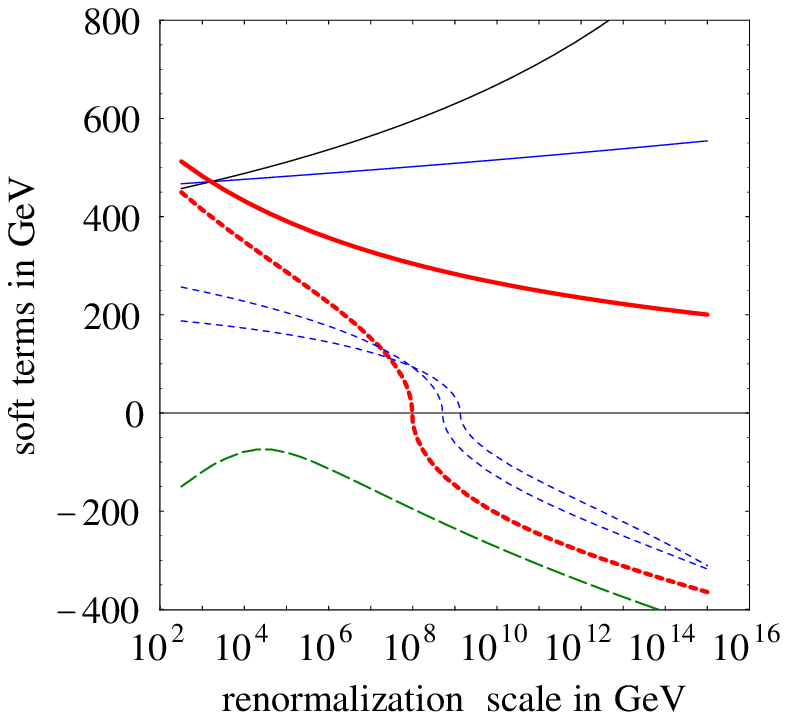}
\putps(8.5,0)(8.5,0){fspettri}{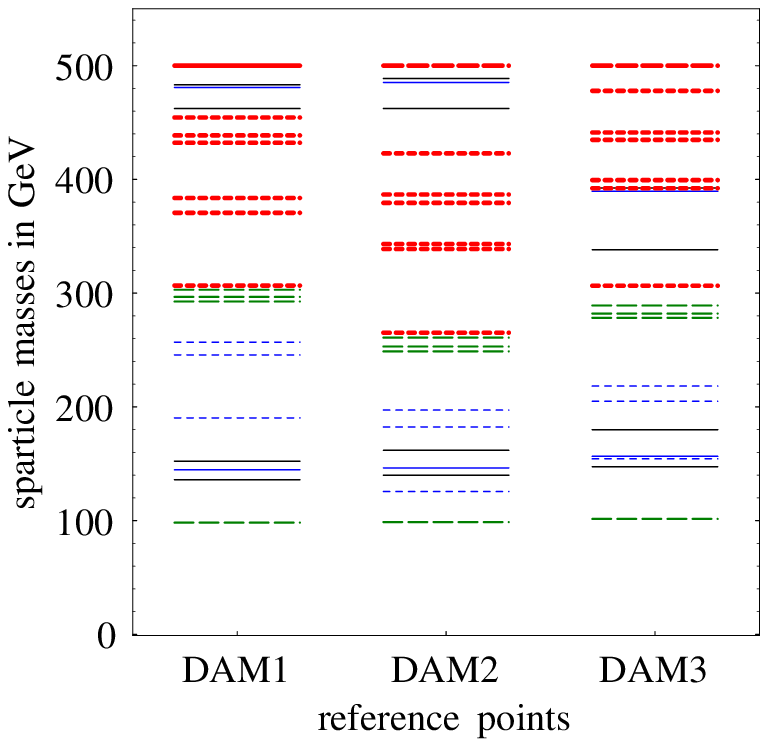}
\end{picture}
\caption[SP]{\em Sample RG evolution of soft terms and qualitatively different sparticle spectra
possible in DAM models.
Notations are explained in the caption of fig.~\ref{fig:spettro}.
\label{fig:sample}}
\end{center}\end{figure*}

However in our  model,  $X$ is a flat direction in 
the supersymmetric limit only lifted by the effects of $F_\phi\not =0$.
The effective action along $X\not =0$ and $\Psi,\bar \Psi=0$
is determined by the running wave function $Z_X(\mu)$ 
\beq
\int d^4\theta~Z_X\big ({\sqrt{XX^\dagger/\phi\phi^\dagger}}\big ) XX^\dagger\, ,
\label{xaction}
\eeq
and gives the effective potential 
\beq
V(X)=m_X^2(|X|)|X|^2\simeq \left |{F_\phi\over 16\pi^2}\right |^2
n\lambda_\Psi^2(X)\left [c_\lambda\lambda_\Psi^2(X)-c_ig_i^2(X)\right ] |X|^2\, ,
\label{runx}
\eeq
where $c_\lambda,c_i>0$, and a sum over the gauge couplings $g_i$ of the 
messengers
is understood. If the running mass $m_X^2$ is positive at large $X$
and crosses zero at some point $X=M_0$,
 the potential has a stable minimum around this point~\cite{cw}.
There exists a choice of parameters for which this happens: the 
positive Yukawa term in eq.~(\ref{runx}) may dominate in the UV 
while the negative gauge contribution may balance it at a lower scale.
For this mechanism to work better one may imagine the presence of a
new and strongly UV free  messenger gauge  interaction. This is because
$SU(3)\times SU(2)\times U(1)$ ends up IR free by the addition
of the messengers. Around the minimum, Re$(X)$ gains a mass $\sim 
(\alpha/4\pi)^3 F_\phi$ which could be of order a few GeV, while
Im$(X)$ is an axion. The crucial result, evident from eq.~(\ref{xaction}),
is $F_X/X=\gamma_X(M_0) F_\phi/2$, a 1-loop quantity instead of the
tree level result $F_X/X=F_\phi$ we mentioned above. 
Therefore, when the messengers are integrated out, their gauge-mediated
contribution to sparticle masses is ${\cal O}(\alpha^2F_\phi)$,
which represents a negligible correction to the original 
${\cal O}(\alpha F_\phi)$ anomaly mediated
 masses. Thus while the gauge beta functions are modified by eliminating
the messengers, the soft terms aren't adjusted to the beta functions
of the low energy theory. Below the scale $M_0$, the RG
flow is {\it deflected} from the AM trajectory. That is why we call
this scenario Deflected Anomaly Mediation (DAM).
Practically the phenomenology of this model is that of 
the MSSM with  boundary conditions for soft
terms at scale $M_0$ given by AM in the MSSM plus $n$ families
of messengers. We give these boundary conditions below. Notice that
the addition of messengers apparently worsens the situation in that it
makes the beta functions more negative. However the gaugino masses are 
also changed: it is the gaugino RG contribution from $M_0$ to $m_Z$
that eliminates all tachyons.
An example of this behaviour for a DAM model with $n=5$ and $M_0=10^{15}\GeV$ is shown in fig.\fig{sample}.

The model is completed by a sector whose dynamics generate $\mu$ and 
$B\mu$. We remind the reader that the generation of these parameters is
 yet another problem of simple AM. As in GMSB, it is quite easy to obtain
the right $\mu$, but it is hard to avoid $B\sim F_\phi\gg m_{\rm weak}$.
These problems are avoided in DAM by considering the addition of one
singlet $S$ coupled via the superpotential
\begin{equation}
\int d^2\theta\left[ 
\lambda_H S \hd \hu+\frac{1}{3}\lambda_S S^3+\frac{1}{2}\lambda_X S^2X\right]\, .
\label{singlet}
\end{equation}
Along $X\not = 0$, the field $S$ is massive and by integrating it out the
following effective operator is generated 
\begin{equation}
\int d^4\theta\, \left \{\hd \hu
 \frac{\lambda_H X^\dagger}{\lambda_X X}\tilde Z\left 
({\sqrt{XX^\dagger/\phi\phi^\dagger}} \right)\,+{\rm h.c.}\right\}\, , 
\label{muterm}
\end{equation}
where $\tilde Z(\mu)$ is the running wave function mixing between $X$ and $S$. 
Eq. \ref{muterm} leads to the following expressions for $\mu$ and $B$
at the scale $M_0$
\beq
\mu=\frac{\lambda_H}{\lambda_X}\left (\gamma_X \tilde Z+\dot{\tilde Z}\right )\frac{F_\phi^*}{2}
\quad\quad\quad B=\frac{\gamma_X^2\tilde Z-\ddot{\tilde Z}}{\gamma_X
\tilde Z+\dot{\tilde Z}}\frac{F_\phi}{2}
\eeq
where the dots represent derivatives with respect to $\ln \mu$.
Both parameters are $\sim \alpha F_\phi\sim m_{\rm weak}$. Notice that
even though the effective operator eq.~(\ref{muterm}) resembles those
of typical GMSB models, $\mu$ and $B$ are the right size since $F_X$
is a 1-loop quantity.


\subsection{Predictions for the soft terms renormalized at $M_0$}
The DAM predictions for the soft terms, renormalized at the high scale $M_0$,
in units of
 $m\equiv F_\phi /(4\pi)^2$, are
\begin{eqnsystem}{sys:m0}
M_i &=&-b_i g_i^2 m\\
A_{RR'R''} &=& (c^R_{i}+c^{R'}_{i}+c^{R''}_{i})g_i^2 m\\
\riga{for any fields $RR'R''$ except}\\[-3mm]
A_t&=&A_{QU\hu }+(\beta_t-\bar\lambda_H^2)m.\\
\riga{The scalar masses of the fields $R$ without significant Yukawa
interactions (sleptons, $d$-squarks and first and second generation of $u$-squarks) are}\\[-3mm]
m_R^2 &=& -b_i c^R_i  g_i^4 m^2.\\
\riga{The soft masses of Higgses and third generation $Q_3$ and $U_3$ squarks
also receive significant
Yukawa contributions}\\[-3mm]
\mhd^2/m^2 &=& -b_i c^L_i  g_i^4 +\delta\\
\mhu^2/m^2 &=& -b_i c^L_i  g_i^4 +\delta+\lambda_t^2 
(-3\beta_t+3\bar\lambda_H^2)\\
m_{U_3}^2/m^2 &=& -b_i c^U_i g_i^4 +\lambda_t^2 (-2\beta_t+2\bar\lambda_H^2)\\
m_{Q_3}^2/m^2 &=& -b_i c^Q_i g_i^4 +\lambda_t^2 (-\beta_t+\bar\lambda_H^2)
\end{eqnsystem}
where all running parameters are renormalized at $M_0$,
$b_i=b_i^{\rm MSSM}+b_i^{\rm mess}=(33/5,1,-3)_i +n$,
the quadratic Casimir coefficients $c_i^R$ are listed in table~\ref{tab:bcude} and
$$\beta_t=(c^Q_i+c^L_i+c^U_i)g_i^2 - 6 \lambda_t^2,\qquad
  \delta = \bar\lambda_H^2(4\bar \lambda_H^2+3\lambda_t^2+\delta'-2c^L_i g_i^2).$$
Finally, $\bar{\lambda}_H$ and $\delta$ are unknown parameters,
related to the unknown parameters in the model Lagrangian as
\bea \nonumber
&&\bar \lambda^2_\Psi=\frac{\lambda^2_\Psi}{Z_{\Psi}Z_{\bar \psi}Z_X (1-|\eta|^2)}\quad\quad
\bar \lambda_X^2=\frac{\lambda_X^2}{Z_S^2Z_X (1-|\eta|^2)^3}\\
&&\bar \lambda_S^2=\frac{\lambda_S^2}{Z_S^3 (1-|\eta|^2)^3}\quad\quad
\bar \lambda_H^2=\frac{\lambda_H^2}{Z_{\hd}Z_{\hu}Z_S (1-|\eta|^2)}\cr
&&\delta'=|\eta|^2(n \bar\lambda_\Psi^2+\frac{5}{2}\bar\lambda_X^2)+
2\bar\lambda_S^2+\bar\lambda_X^2-(\bar\lambda_S \bar\lambda_X^*\eta^*+{\rm h.c.}). \nonumber
\eea
where $\eta={\tilde Z}/{\sqrt{Z_XZ_S}}$.
Notice that $|\eta|<1$ is required
for the model to be stable (positive kinetic terms).
Then $\delta'$ is positive definite and $\delta$ is positive. We will see 
below that these
extra positive contributions to the Higgs mass parameter, together
with the requirement of correct electroweak symmetry breaking
(EWSB), lead to an upper bound on $\mu/m$.

As is often the case, the model-dependent couplings introduced to generate
the $\mu$ and $B\mu$ terms also affect the Higgs mass parameters.
In this concrete model they also affect the soft parameters of the third generation squarks.
Since 4 soft masses depend on only two unknown parameters ($\bar{\lambda}_H$ and $\delta$)
there are testable predictions.
On the contrary the $\mu$ and the $B\mu$ terms
are determined by more than two additional unknown combinations of parameters;
therefore, we consider them as free parameters and
do not give their explicit expression in terms of model parameters.
Even assuming real Yukawa couplings in the messenger sector,
the observable sign of the $B\mu$ term is not predicted.
However, if for some reason the kinetic mixing term $\tilde{Z}$ is small,
CP phases can be rotated away. 
The model then predicts the sign of  $B\mu$ and gives
one relation between $\mu$, $B\mu$ and the soft terms.

We have here neglected the effects of the other Yukawa couplings,
including the possibly significantly $\tau$ and $b$ ones.
If $\tan\beta$ is large their effect should be added.
They should also be taken into account
when studying the predictions for `fine details' of the spectrum
(like the mass splitting between 
$\tilde \tau_1\simeq \tilde{\tau}_R$ and $\tilde{e}_R,\tilde{\mu}_R$
and the $q/\tilde{q}$ mixing angles at the gaugino vertices induced by the CKM matrix).

The soft terms at the electroweak scale are obtained by renormalizing
their values at $M_0$ listed in this section
with the usual MSSM RG equations.
The standard semi-analytic solutions
cannot be applied in this case since gaugino masses do not obey unification relations, $M_i\propto \alpha_i$.
In appendix A we write the RG evolution for the soft terms starting 
from the most general 
boundary conditions
in terms of a minimal number of `semi-analytic' functions.
DAM models predict $M_i\propto (b_i^{\rm MSSM}+n)\alpha_i$.
In this particular case the semi-analytic solutions could
 be further simplified.

\begin{figure*}[t]
\begin{center}
\begin{picture}(17.7,6)
\putps(0.5,0)(0.5,0){fparam0}{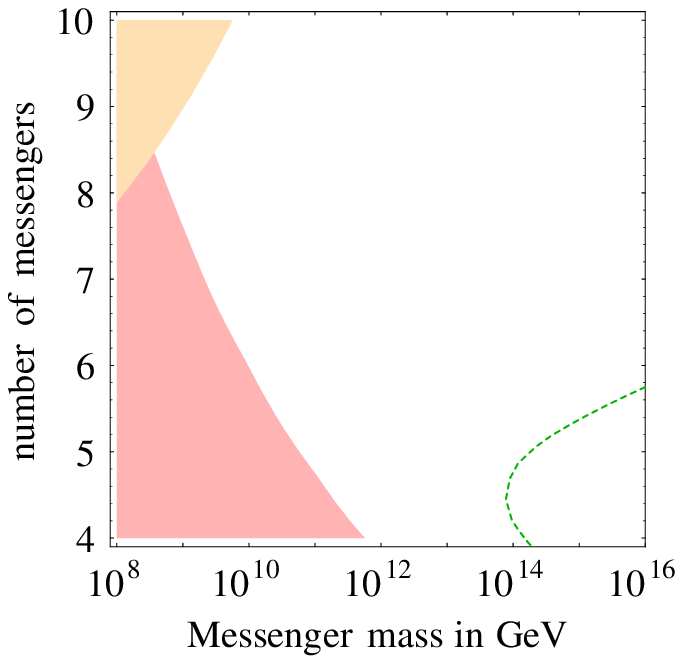}
\putps(9.0,0)(9.0,0){fparam1}{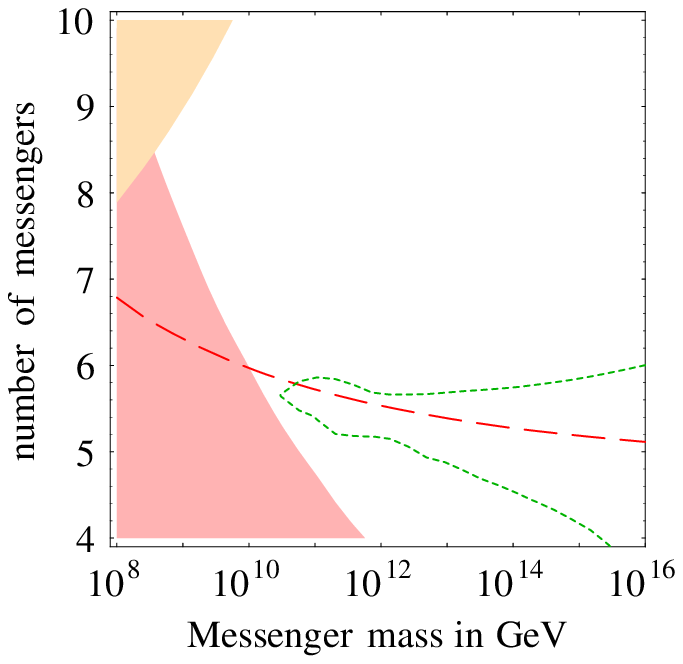}
\put(4.00,7){fig.\fig{allowed}a}
\put(12.2,7){fig.\fig{allowed}b}
\end{picture}
\caption[SP]{\em Allowed values of the main unknown model parameters, $n$ and $M_0$
for $\tan\beta=4$, $\lambda_t(\MGUT)=0.5$, and small ($\bar{\lambda}_H=0$, fig.\fig{allowed}a)
or significant ($\bar\lambda_H=1$, fig.\fig{allowed}a) Yukawa messengers.
In the unshaded regions of the $(n,M_0)$ plane
tachyonic sleptons are avoided without too many light messengers.
Below the dashed line $\mhu^2$ is positive, so that EWSB
is possible only with appropriate correlation between the parameters.
Inside (outside) the dotted lines the 
lightest superpartner is a higgsino (almost always a slepton).
\label{fig:allowed}}
\end{center}\end{figure*}


\section{The sparticle spectrum}
The predictions for the soft terms depend on 7 parameters.
The gaugino masses  
depend only on $n$ (the messenger contribution to the gauge $\beta$ functions);
soft terms of first and second generation sfermions depend only on $n$ and $M_0$ (the messenger mass);
while
soft terms of third generation sfermions and higgses depend
also on the (imprecisely known) top Yukawa coupling at $M_0$,
and on the possible messenger couplings $\bar\lambda_H$ and $\delta$.
The $\mu$ and $B\mu$ terms can be considered as free parameters,
and are fixed in our analysis by the conditions of successful EWSB.

The dependence on $\lambda_t$
is stronger than in gauge mediation or supergravity models.
The unknown parameters $\bar\lambda_H$ and $\delta$ can give important corrections
when $n$ is not too large ($n\circa{<}10$): in these cases they always increase the value of
$\mhu^2(Q)$, and thus reduce the value of $\mu$ that gives a correct EWSB.

Even if all parameters are important, $M_0$ and $n$ are the ones that
control most of the sparticle spectrum (the gauginos and the sfermions).
In fig.\fig{allowed} we show the phenomenologically acceptable
range of $(M_0,n)$ for $\lambda_t(\MGUT)=0.5$
and small messenger couplings.
Shaded regions are excluded because the gauge couplings run to infinity before the unification scale
(if $n$ is too large), or
because one slepton is tachyonic (if $n$ is too low).
If $n<4$ there are tachyonic sleptons, as in pure AM where $n=0$.
If $n=4$ sleptons can have positive squared masses, but also $\mhu^2$ is positive.
When $n>4$ it is possible to have negative $\mhu^2$ and positive sfermion masses
unless $M_0$ is too low.
In all the parameter space there exist unphysical deeper minima
(since $m_{\tilde{\ell}}^2<0$ at high field values, see fig.\fig{sample}).
There is no reason for excluding the model for this reason.
Quantum and thermal tunneling rates are negligible~\cite{CCBcosm1}.
Moreover within standard cosmology there exist plausible mechanisms~\cite{CCBcosm2}
that naturally single out the desired physical minimum closer to the origin.
A possible source of cosmological
problems is the modulus $X$, since its mass cannot exceed a few
GeV.  Therefore to avoid large modulus fluctuations
we must assume $X$ to be already around its minimum when
the temperature of the universe is somewhat below $\langle X\rangle$.
Then, since $X$ is only coupled to the MSSM by non-renormalizable interactions
at low energy, thermal fluctuations will not affect it.


\begin{figure*}[t]
\begin{center}
\begin{picture}(17.7,6)
\putps(0.5,0)(0.5,0){fFTn5}{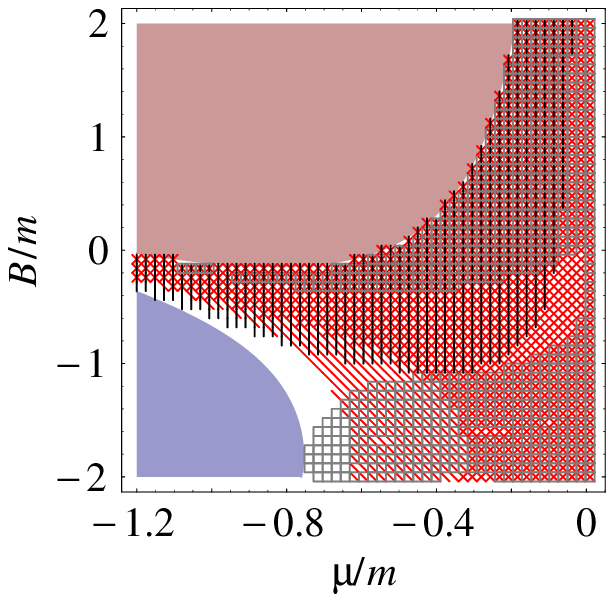}
\putps(9,0)(9,0){FTsugra}{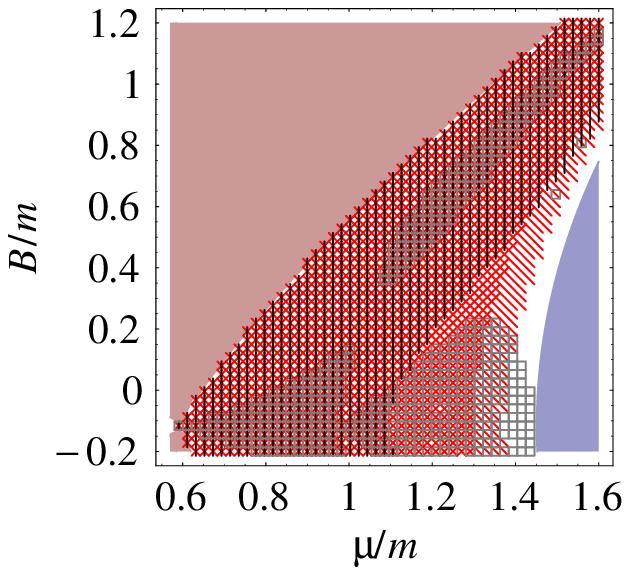}
\put(4.00,6.5){fig.\fig{nat}a}
\put(12.2,6.5){fig.\fig{nat}b}
\end{picture}
\caption[SP]{\em 
The size of the allowed regions (empty  regions)  of the parameter space $(\mu/m,B/m)$
indicates how `natural' is the model.
Fig.\fig{nat}a refers to our reference `DAM1' model
while in fig.\fig{nat}b we show for comparison a mSUGRA model with $m_0=m_{1/2}$ and $A_0=0$.
The shaded regions are excluded because correct EWSB is not possible, while
regions marked with different symbols are now experimentally excluded (see text).
\label{fig:nat}}
\end{center}\end{figure*}

\subsection{EWSB and naturalness}
In most of the acceptable parameter space only the higgs field $\hu$ has a negative squared mass term, $\mhu^2<0$,
so that EWSB is induced by supersymmetry breaking in the usual way.
However, $\mhu^2$ is positive for certain values of the parameters:
this happens for $n=4$ (unless $\lambda_t$ and $\bar{\lambda}_H$ are small);
it also happens for higher values of $n$ below the dashed lines in fig\fig{allowed}
if $\bar{\lambda}_H\sim 1$.
With a positive $\mhu^2$ it is still possible to break electroweak
symmetry, but only
in the narrow region of the parameter space where the $\mu$ and $B\mu$ 
terms give appropriate mixings in the higgs mass matrix.
Moreover this situation tends to give values of $\tan\beta$ close to 1,
so that the lightest higgs mass is below its experimental bound unless the sparticles are very heavy.
For these reasons we do not consider this possibility attractive, and we
 will restrict our analysis to the more interesting case $n\ge 5$.

Strong, non-preliminary constraints on the parameter space are now given by
LEP and Tevatron experiments.
The bounds $m_{\chi^\pm} \circa{>}90\GeV$, 
$m_h\circa{>}85\GeV$ and $M_3\circa{>}(180\div 250)\GeV$
are satisfied only in
a small portion of the parameter space of `conventional' supersymmetric models
(like mSUGRA and GMSB), implying that the 
EWSB scale is unexpectedly smaller than the unobserved sparticle masses.
How unnatural this situation is in any given model depends on two different
 characteristics of the model:
\begin{enumerate}
\item How light is the $Z$ boson mass with respect to the soft terms?
Since EWSB is induced by supersymmetry breaking,
$M_Z^2$ is predicted to be a sum  of various squared soft mass terms 
(often dominated by the gluino contribution).

\item How strong are the bounds on model parameters induced by the experimental
bounds on sparticle masses?
The naturalness problem becomes more stringent
in the presence of an indirect bound on $M_3$ stronger than the direct Tevatron bound on $M_3$.

\end{enumerate}
Concerning the second point, in SUGRA and GMSB gaugino masses obey unification relations
so that the LEP bound on the chargino mass gives an indirect bound on the gluino mass,
$M_3\circa{>}300\GeV$, somewhat stronger than the direct Tevatron bound, 
$M_3\circa{>}220\GeV$ 
(valid if $m_{\tilde{q}}\approx M_3$, as in our model).
This undesired feature is not present in the scenario under study,
basically for all appealing values of the parameters.
However, as it happens in GMSB, the bound on the selectron mass
gives an indirect bound on $M_3$ which is stronger than the Tevatron bound.
In conclusion, for what concerns point 2, DAM is not better than 
`conventional' models.


On the contrary DAM makes a somewhat more favourable prediction regarding 
point 1.
It predicts a cancellation in the EWSB conditions for $M_Z^2$, because 
the positive radiative $\Ord(M_3^2)$ contribution
to $M_Z^2$ is partially canceled by negative radiative $\Ord(m_{\tilde{q}}^2)$ contributions
(in DAM models all sfermion squared masses are negative, before including RG
 corrections).


Putting it all together, DAM models suffer from some naturalness problem.
This is mainly because the experimental bounds on sparticle
masses are satisfied only in a small region of parameter space \cite{nat}.
This is shown in fig.\fig{nat}a, where we display the allowed portion of 
the parameter space
for fixed $n=5$, $M_0=10^{15}\GeV$ and $\lambda_t(\MGUT)=0.5$
and assuming that the Yukawa couplings of the messengers are negligible.
With this assumption the soft terms only depend on 3 parameters:
$m$ (the overall scale of anomaly mediated soft terms), the $\mu$-term and $B$.
The EWSB condition allows to compute the overall SUSY scale $m$ and 
$\tan\beta$ in terms of two dimensionless ratios
($\mu/m$ and $B/m$ in figs.\fig{nat}, all renormalized at $M_0$).

In fig.\fig{nat} we have shaded the regions where correct EWSB is not possible,
and marked with different symbols the points of the parameter space where
some sparticle is too light.
Sampling points marked with a
(\setlength{\unitlength}{1em}%
\begin{picture}(1,1)(0,0)\put(-0.2,1){\line(1,-1){1}}\end{picture},
\begin{picture}(1,1)(0,0)\put(0,0){\line(1,1){1}}\end{picture},
\begin{picture}(1,1)(0,0)\put(0,0){\line(0,1){1}}\end{picture},
\begin{picture}(1,1)(0,0)\put(0,0.3){\line(1,0){1}}\end{picture})
are experimentally excluded because
a (gluino, chargino, selectron, higgs) is too light.
Regions where ${\rm BR}(B\to X_s\gamma)$ differs from its SM value by more than $50\%$ are marked with a
\begin{picture}(1,1)(0,0)
\put(0,0){\line(1,0){1}}
\put(0,1){\line(1,0){1}}
\put(0,0){\line(0,1){1}}
\put(1,0){\line(0,1){1}}\end{picture}.\setlength{\unitlength}{1cm}
We have restricted our plots to signs of $\mu$ and $B$ such that
the interference between charged higgs and chargino contributions to the $b\to s\gamma$ decay amplitude is destructive.
With a constructive interference
the indirect bounds on sparticle masses from ${\rm BR}(B\to X_s\gamma)$ are
stronger than the direct accelerator bounds and restrict the allowed
parameter space to a very small region, smaller than our resolution of 
figs.\fig{nat}.

We see that different portions of the parameter space are excluded by
different combination of the bounds on gluino, charged higgsino, slepton and higgs masses.
Since DAM models look somewhat disfavoured by naturalness considerations\footnote{In DAM models
with $n\sim 5$ the fine tuning (FT~\cite{FT}) of $M_Z$ with respect to 
the soft terms is typically low.
By choosing appropriate values of the unknown Yukawa couplings it is even possible to get FT $\approx 1$.
However this does not mean that DAM models are perfectly natural:
since the soft terms depend on unknown Yukawa couplings
the FT with respect to just the soft terms is not an adequate 
measure of naturalness~\cite{FTcoupling}.},
we also show in fig.\fig{nat}b that a typical mSUGRA model
(assuming $A_0=0$ and $m_0=m_{1/2}$ in order to make a plot in the 
$(\mu/m_0, B/m_0)$ plane)
has similar problems. Moreover, also gauge mediated models have a naturalness
 problem, mainly because they predict light right-handed sleptons. As for
pure AM ($n=0$), it predicts tachyonic sleptons, and it
also has some naturalness problem:
a chargino heavier than the LEP2 kinematical reach limit, $M_2\circa{>}M_Z$,
would imply that the contribution from $\mhu^2$ to $M_Z^2$ is $\sim 100$ times larger than $M_Z^2$ itself.
Adding a universal contribution to scalar masses~\cite{RS,GGW,FM} 
eliminates the tachyons but does not improve naturalness.
On the other hand, DAM models also do better on the problem of naturalness.

\begin{figure*}[t]
\begin{center}
\begin{picture}(17.7,8)
\putps(0,0)(0,0){fs15}{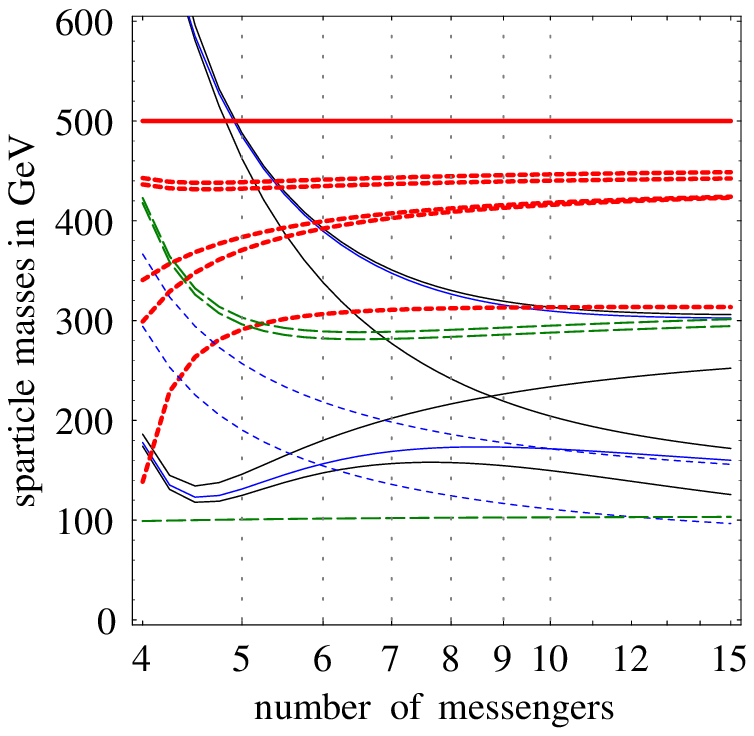}
\putps(8.9,0)(8.9,0){fsM}{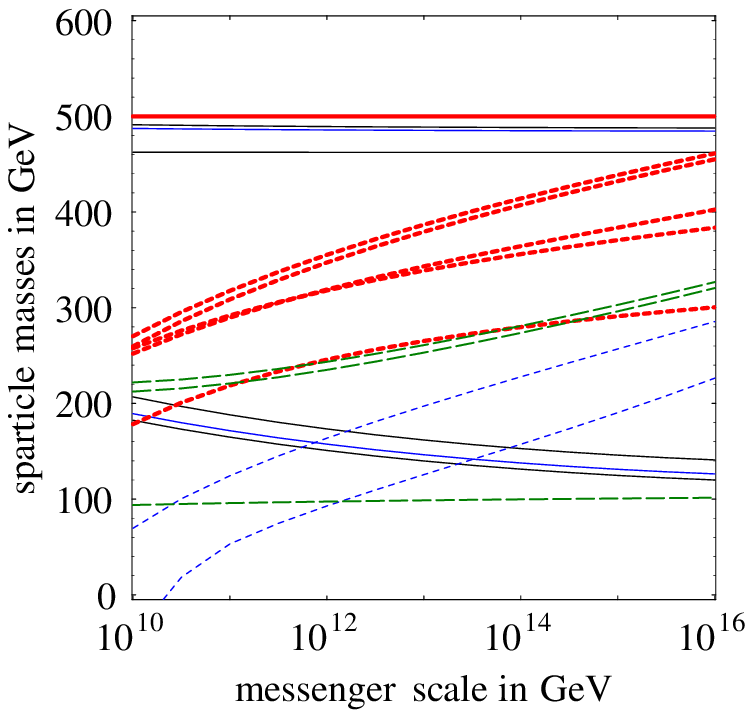}
\put(16.3,6.5){$\tilde{g}$}
\put(16.3,6.1){$\tilde{W}$}
\put(16.3,5.7){$\tilde{B}$}
\put(16.3,5.2){$\tilde{q}$}
\put(16.3,4.6){$h_\pm$}
\put(16.3,4.1){$\tilde{\ell}_L$}
\put(16.3,3.5){$\tilde{\ell}_R$}
\put(16.3,2.6){$\tilde{h}$}
\put(16.3,2.2){$h$}
\put(4.00,7.8){fig.\fig{spettro}a}
\put(12.6,7.9){fig.\fig{spettro}b}
\end{picture}
\caption[SP]{\em Spectrum of sparticles as function of $n$ for $M_0=10^{15}\GeV$ (fig.\fig{spettro}a)
and as function of $M_0$ for $n=5$ (fig.\fig{spettro}b)
at fixed $M_3=500\GeV$, $\lambda_t(M)=0.5$ and negligible Yukawa messengers.
Dashed (long dashed, continuous) lines refer to sfermions (higgses, fermions).
Thin (thick and red) lines refer to uncoloured (coloured) sparticles.
Black (blue) lines refer to the neutralinos (charginos and sleptons).
\label{fig:spettro}}
\end{center}\end{figure*}

\subsection{The sparticle spectrum}
We now continue our analysis studying the spectrum of sparticles in the
 allowed portion of the parameter space. 

Before going on,
we must anticipate (see the discussion in section 6) that the
 LSP of our models is the fermionic component of the $X$ modulus.
This fact is important as it allows a charged
NSLP (sometimes a slepton). However,
over the parameter space allowed in fig. 2, the NLSP decays into LSP
always outside the detector. Therefore, the NLSP is practically a
 stable particle  and the LSP plays
no role in collider phenomenology.

In fig.\fig{spettro} we plot the spectrum
as a function of $n$ for $M_0=10^{15}\GeV$ (fig.\fig{spettro}a) and as a 
function of $M_0$ for $n=5$ (fig.\fig{spettro}b).
In both cases we have assumed $M_3=500\GeV$, $\tan\beta=4$, 
$\lambda_t(\MGUT)=0.5$ and negligible messenger Yukawa couplings 
and  computed the $\mu$ term from the condition of correct EWSB.
Although no unique pattern emerges over all the parameter space, 
we try to summarize the main features of the spectrum
in the following way:
\begin{enumerate}

\item[0.] 
The NLSP is usually a slepton or a neutral higgsino.
The mass splitting between sleptons receives three different computable
contributions; all of them (apart from a less important
RG effect) tend to make
an almost right-handed $\tilde{\tau}$ state the lightest slepton.
Although the $\tilde{\tau}_R$ is often lighter than the higgsino (see fig.\fig{allowed}),
it is always possible to force an higgsino NLSP by
increasing the value of the unknown messenger Yukawas 
which decreases the value of $\mu$ that gives the correct EWSB.
When $n=4$ the NLSP can be a stop, while for large $n\circa{>}10$ the NLSP can be a bino.


\item[1.] When $n=5$ the NLSP is most often a neutral 
higgsino, sleptons are light,
and all gauginos have a comparable mass {\em above\/} the squark masses.

\item[2.] When $n=6,7,8$ the electroweak gauginos are lighter than the 
squarks, but heavier than the higgsinos.

\item[3.] 
When $n\gg 1$ the sfermion and gaugino masses
are dominated by the pure anomaly mediated contribution to gaugino masses.

\end{enumerate}
As discussed in the next section, features 1 and 2 listed above 
give characteristic manifestations
at hadronic colliders.
It is more difficult to distinguish DAM models with larger $n$ from
mSUGRA or GMSB at hadron colliders, even if for quite 
large values of $n$  the mass spectrum
remains significantly different from the one with unified gaugino masses.
For example if $n=20$ the ratio $M_1/M_3$
(connected in a simple way to the measurable ratio between the bino and the gluino masses)
is still $50\%$ higher than in the `unified gaugino' case.

In the following section we perform more detailed studies by selecting 
three reference points in the DAM parameter space 
that capture the main characteristics of the model:
\begin{description}

\item[DAM1:] we choose
$n=5$, $M_0=10^{15}\GeV$, $\bar\lambda_H=0$, $\lambda_t(\MGUT)=0.5$, $M_3=500\GeV$
in order to have a characteristic DAM model with $n=5$ and higgsino NLSP.

\item[DAM2:] we choose
$n=6$, $M_0=10^{15}\GeV$, $\bar\lambda_H=0$, $\lambda_t(\MGUT)=0.5$, $M_3=500\GeV$
in order to have a characteristic DAM model with $n=6$ and slepton NLSP.

\item[DAM3:] we choose $n=6$, $M_0=10^{15}\GeV$, $\bar\lambda_H=1$, $\lambda_t(\MGUT)=0.5$, $M_3=500\GeV$.
DAM3 is similar to DAM2, except the NLSP is a neutral higgsino.

\end{description}
The spectra corresponding to these three
sets of parameters are shown in fig.\fig{sample} and listed in 
tables~\ref{mtable1} and \ref{mtable2}.
Using these three examples we will now
illustrate the phenomenology at high-energy colliders.


\section{Signals at collider}

The experimental manifestation of supersymmetry at hadron
colliders like the Tevatron and 
the LHC depends strongly on how the supersymmetric particles
are ordered in mass, and on the nature of the lightest superpartner
of ordinary particles 
(stable/unstable, charged/neutral).  The model under study has
strong dependences on the parameters of the theory, and therefore
does not make unique predictions for these important issues relevant
to collider physics.  Furthermore, measuring the parameters at
a high-energy hadron collider is not a straightforward task.  Nevertheless,
we would like to point out  some expectations for these
models at hadron colliders despite the above difficulties.


\begin{table}[t]
\begin{center}
\begin{tabular}{cccc} 
\multicolumn{4}{c}{Sparticle spectrum in DAM model 1}\\
\hline \hline
Sparticle  & mass \qquad &Sparticle & mass \\ \hline
$\wtil g$  		& 500 	& 			&	\\
$\wtil \chi_1^\pm$	& 145	& $\wtil \chi_2^\pm$	& 481	\\
NLSP={$\wtil \chi_1^0$}	& 136	& $\wtil \chi_2^0$	& 152	\\
$\wtil \chi_3^0$	& 462	& $\wtil \chi_4^0$	& 483	\\
$\wtil u_L$		& 432	& $\wtil u_R$		& 384	\\
$\wtil d_L$		& 439	& $\wtil d_R$		& 371	\\
$\wtil t_1$		& 306	& $\wtil t_2$		& 454	\\
$\wtil b_1$		& 371	& $\wtil b_2$		& 406	\\
$\wtil e_L$		& 257	& $\wtil e_R$		& 190	\\
$\wtil \nu_e$		& 246 	& $\wtil \nu_\tau$	& 246	\\
$\wtil \tau_1$	& 190	& $\wtil \tau_2$	& 257	\\
$h^0$			& 98    & $H^0$			& 297	\\
$A^0$			& 293	& $H^\pm$		& 303	\\
\hline \hline
\end{tabular}
\end{center}
\caption{\em Masses of the SUSY particles, in GeV, for the DAM model point 1.
\label{mtable1}}
\end{table}

\begin{table}[t]
\begin{center}
\begin{tabular}{cccc|cccc}
\multicolumn{4}{c}{Sparticle spectrum in DAM model 2}&
\multicolumn{4}{c}{Sparticle spectrum in DAM model 3}\\
\hline \hline
Sparticle  & mass \qquad &Sparticle & mass &
Sparticle  & mass \qquad &Sparticle & mass \\ \hline
$\wtil g$  		& 500 	& 			&	& $\wtil g$  		& 500 	& 			&	\\
$\wtil \chi_1^\pm$	& 176 	& $\wtil \chi_2^\pm$	& 381	& $\wtil \chi_1^\pm$	& 151	& $\wtil \chi_2^\pm$	& 381	\\
$\wtil \chi_1^0$	& 165	& $\wtil \chi_2^0$	& 187	& NLSP={$\wtil \chi_1^0$}& 141	& $\wtil \chi_2^0$	& 162	\\
$\wtil \chi_3^0$	& 337	& $\wtil \chi_4^0$	& 382	& $\wtil \chi_3^0$	& 337	& $\wtil \chi_4^0$	& 382	\\
$\wtil u_L$		& 435	& $\wtil u_R$		& 399	& $\wtil u_L$		& 435	& $\wtil u_R$		& 399	\\
$\wtil d_L$		& 441	& $\wtil d_R$		& 392	& $\wtil d_L$		& 441	& $\wtil d_R$		& 392	\\
$\wtil t_1$		& 326	& $\wtil t_2$		& 465	& $\wtil t_1$		& 313	& $\wtil t_2$		& 470	\\
$\wtil b_1$		& 392	& $\wtil b_2$		& 412	& $\wtil b_1$		& 392	& $\wtil b_2$		& 410	\\
$\wtil e_L$		& 218	& coNLSP=$\wtil e_R$ 	& 154	& $\wtil e_L$		& 218	& $\wtil e_R$		& 154	\\
$\wtil \nu_e$		& 205 	& $\wtil \nu_\tau$	& 205	& $\wtil \nu_e$		& 205 	& $\wtil \nu_\tau$	& 205	\\
coNLSP=$\wtil \tau_1$	& 154	& $\wtil \tau_2$	& 218	& $\wtil \tau_1$	& 154	& $\wtil \tau_2$	& 218	\\
$h^0$			& 99	& $H^0$			& 283	& $h^0$			& 101   & $H^0$			& 290	\\
$A^0$			& 278	& $H^\pm$		& 289	& $A^0$			& 286	& $H^\pm$		& 296	\\
\hline \hline
\end{tabular}
\end{center}
\caption{\em Masses of the SUSY particles, in GeV, for the DAM model
point 2 (left columns)
and for DAM model point 3 (right columns).
\label{mtable2}}
\end{table}

The most important feature of the model we are presenting here is
the relatively small mass gap between all the gauginos.  One immediate
consequence of this is a changed interpretation of gluino mass bounds from
LEP2 results.  The $e^+e^-$ LEP2 collider does not produce gluinos directly, 
yet it does probe the Winos very effectively.  Limits on the charged
Wino mass from the four LEP collaborations
are nearly $100\gev$~\cite{lepchargino}, the exact value
depending on the details of the 
full supersymmetric spectrum.  This can be interpreted as a limit
on the gluino mass of about $m_{\tilde g}\gsim 300\gev$, provided 
we assume gaugino mass unification.  Therefore, if the Tevatron
finds a gluino with mass less than $300\gev$, by any of the known
discovery channels, that would be
one piece of evidence for the AM models.  Current direct
limits on the gluino 
mass are approximately $m_{\tilde g}\gsim 185\gev$ in R-parity conserving supersymmetric
models with $m_{\tilde q}\gg m_{\tilde g}$, and 
$m_{\tilde g}\gsim 220\gev$ when
$m_{\tilde q}=m_{\tilde g}$~\cite{tevatrongluino}.  

To be convinced that the DAM model is correct, much
additional evidence must be gathered consistent with the model.
Useful
observables at hadron colliders include total rates above background in
large lepton/jet multiplicity events with missing energy, 
invariant mass peaks of decaying heavy particles, kinematic edges to
lepton or jet invariant mass spectra, and exotic signatures such as
a highly ionizing track associated with a stable, heavy, charged
particle track passing through the detector.  All of these methods
can be used to uncover evidence for supersymmetry and to help determine
precisely what model is being discovered.

DAM models have several gross features that may be keys to distinguishing
them from other models, such as mSUGRA and minimal GMSB. 
One such feature that we mentioned above is
the relatively small mass difference between all the sparticles in the
spectrum.  Typical parameter choices in models of
mSUGRA and especially GMSB have nearly an order of magnitude
difference between the lightest supersymmetric partner (not counting the
gravitino) and the heaviest partner.  The heaviest of these sparticles
are usually the strongly interacting squarks and gluinos.  
Consequently, unless sparticles are much heavier than the top quark, in DAM models the decays $\tilde{g}\to \tilde{t}_2 t$
and $\tilde{t}_1 \to t N$ are usually kinematically forbidden
($\tilde{t}_1$ is the lighter stop and $\tilde{t}_2$ is the heavier stop).
Therefore in DAM models it is not unusual to have at most two top quarks per event, while four top quarks can be present in mSUGRA and GMSB models.

\medskip

\begin{table*}[t]
\label{lepton multiplicity}
\centering
\begin{tabular}{cccc}
\hline\hline\\[-3mm]
\parbox{2cm}{\centering Number of \\ leptons} &  DAM Model 1   & 
\parbox{4cm}{\centering DAM Model 1 \\ with $M_i=\alpha_iM_3/\alpha_s$}& DAM Model 3\\[2mm]
\hline
0    & 813 (741)  & 714 (700)  & 161 (122)  \\
1    & 85 (129)  & 105 (117) &  169 (137) \\
2    & 24 (48)  &  12 (13) &  233 (248) \\
3    & 1 (5)  &  1 (2) &  99 (117) \\
4    & 0 (0)   & 1 (1)  & 57 (84)  \\ 
5    & 0 (0)   & 0 (0)  &  9 (17) \\
6    & 0 (0)   & 0 (0)  &  2 (5) \\
$7+$ & 0 (0)   & 0 (0)  &  0 (0) \\
\hline\hline
\end{tabular}
\caption{\em From DAM model point 1, the
lepton multiplicity in 1000 simulated LHC events with 
at least $200\gev$ of total missing energy.  Leptons are counted if
they have $\eta<3$ and $p_T>10$~GeV. The numbers in parenthesis
have no $p_T$ cut on the leptons.  In the third column the
spectrum is the same as DAM model point 1, except the $M_1$ and
$M_2$ masses are GUT normalized.  In the last column, the lepton
multiplicity is given for DAM model point 3, which has significant
production of leptons due to on-shell cascades 
of $\tilde q\to \tilde W \to \tilde l$.}
\end{table*}

\subsection{Total rates with two stable sleptons}

Over much of the DAM parameter space the lightest supersymmetric partner to
be produced in the detector is the $\tilde l_R$.  For example,
analyzing DAM model point 2 (see Table~\ref{mtable2}), 
we find that the NLSP is $m_{\tilde l_R}=155\gev$ and
$M_1,M_2,M_3,\mu$ are $334,364, 500$ and $-176\gev$ respectively.  Production
of gauginos, squarks and sleptons all end up producing the lightest
state $\tilde l_R$, which can be discovered rather easily by the
detectors.  The total supersymmetry production rate at the Tevatron
with $\sqrt{s}=2\tev$ is more than 200~fb, and with several fb${}^{-1}$
expected at Tevatron runII, this choice of parameters for the model
would be detected, despite superpartners not being 
kinematically accessible at LEP2.  A
careful analysis of run I data may even be able to discover or
definitively rule out the parameter choices made for this example.

GMSB is another model that has a large parameter 
space for (quasi)-stable sleptons.  If stable, charged tracks are discovered
at the Tevatron, the first task will be to find the mass of the particle,
and then determine the rest of the spectrum that gave rise to this sparticle.
Finding the mass is relatively straightforward once there is a significant
signal.  Timing information along with $dE/dx$ measurements
as the particle passes through
the detector are useful in this regard.  Determining what model these stable
tracks come from is much more difficult.  One beginning step will be to
estimate total supersymmetry production rate based on all 
$\sigma(\tilde l_R\tilde l_R+X)$ signatures.  This can then be
compared between the DAM model presented here and, say, minimal GMSB.

If we apply the slepton and chargino mass limits from LEP2 to GMSB, and
then analyze expectations for the Tevatron, we find that squark and gluino
production are not significant in supersymmetry searches at the Tevatron.
This is even true when the $\tilde l_R$ is the NLSP and does not decay
in the detector -- perhaps the most likely possibility~\cite{dtw} in GMSB with
$N_{5+\bar 5}\geq 2$.
Neglecting potentially important detector efficiency issues, every
event that produces superpartners will be registered and tagged as
a supersymmetry event since stable sleptons yield such an exotic 
signature in the detector~\cite{fengmoroi,Martin:1998vb}.  
Production of sleptons, gauginos, higgsinos,
and squarks all will decay ultimately to two charged sleptons plus 
standard model particles.  Therefore, we can speak about the discovery 
of these models solely by analyzing the two sleptons and ignoring all
other associated particles in the events, just as we did for the DAM.
In this case, there is very little variability in the total rate
for $\tilde l_R\tilde l_R+X$, and the rate depends mostly on the number of
$5+\bar 5$ messengers.  In Fig.~\ref{sigmaSUSY} 
we plot the range allowed for total
supersymmetry production~\cite{ISAJET} 
in GMSB with moderate to small $\tan\beta$
as a function of $m_{\tilde e_R}$.
(Distinguishing between low and high $\tan\beta$ can be accomplished by
careful analysis of the associated particles in $X$~\cite{Martin:1998vb}.)  
The upper line
corresponds to $N_{5+\bar 5}=2$ and the lower line corresponds
to $N_{5+\bar 5}=\infty$.  In contrast, recall from the paragraph above
that a typical DAM set of parameters yielded a total cross-section
of 200~fb for $m_{\tilde l_R}=155\gev$ because the squarks and gluinos
are much lighter and contribute to the signal.  Therefore, a first step
in distinguishing between DAM models and GMSB models is to measure
$m_{\tilde e_R}$ directly from stable, charged particle track analysis,
and then compare the total measured rate of $\sigma (\tilde l_R\tilde l_R+X)$
to Fig.~\ref{sigmaSUSY}.
\jfig{sigmaSUSY}{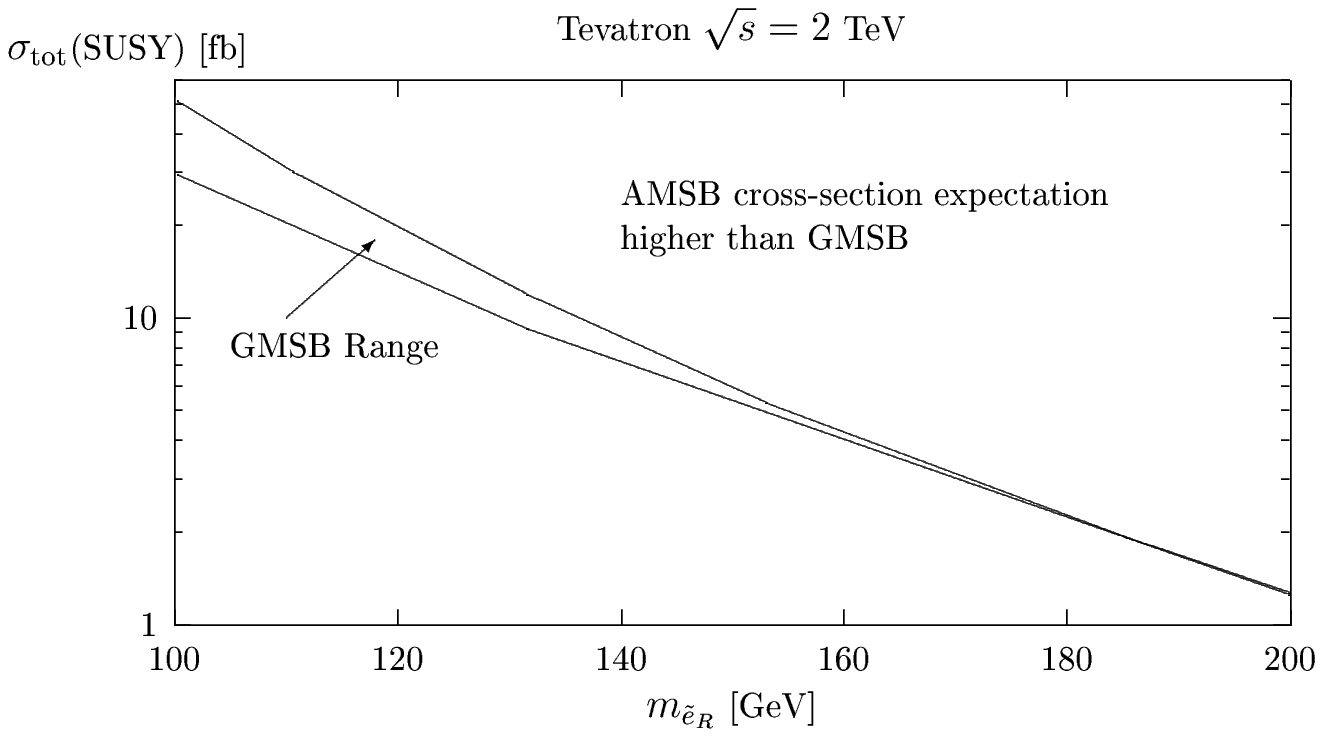}{Total cross-section for 
supersymmetry production at the Tevatron.  The upper line is for GMSB
with $N_{5+\bar 5}=2$ messengers, and the lower line is for 
$N_{5+\bar 5}=\infty$ messengers.  Minimal
GMSB models are expected to fall within
these two lines.  DAM models, by contrast, are expected to be have 
much higher cross-sections since squark and gluinos masses are generally much
lighter for the same $m_{\tilde e_R}$.}
\bigskip


\subsection{Lepton multiplicity and $p_T$ distributions}

Other important observables in supersymmetric events are the lepton 
multiplicity and $p_T$ distributions.  These are often sensitive to the
mass hierarchies in the supersymmetric model.  For example, a large
source of high $p_T$ leptons in mSUGRA models is the cascade decays
through $\chi^\pm_1\to l\nu\chi^0_1$.  The mass difference between
$m_{\chi^\pm_1}$ and $m_{\chi^0_1}$ is large, and the $\chi^\pm_1$
state is expected to participate significantly in the cascade decays of 
the heavier squarks and gluino down to the LSP.  

In contrast, the DAM model has relatively
fewer sources of high $p_T$ leptons because of the near degeneracy
of the NLSP and the next least massive chargino and neutralino.
For example,
in DAM model point 1 (see Table~\ref{mtable1}), we find
the quasi-stable NLSP is $\chi_1^0\sim \tilde H$, 
$M_1,M_2,M_3=461,468,500$~GeV and $380\gev\lsim m_{\tilde q}\lsim 440\gev$.
Production of gluinos and squarks, while large in this model, more rarely
produce sleptons because $m_{\tilde q}\lsim M_1,M_2$.  Instead,
$\tilde q$ like to decay directly to a quark and a Higgsino
with no intermediate leptons in a cascade decay.  Leptons
can arise however from $\chi^-_1\to l^-\nu\chi^0_1$, but these leptons
are somewhat softer because of the near degeneracy between the mostly
Higgsino $\chi^-_1$ and $\chi^0_1$ states.  In the particular example
given here, the mass splitting between the
lightest chargino and the lightest neutralino is about $10\gev$.
The other significant
source of leptons comes from third family superpartner production
and decay. Since the stop and sbottom squarks are rather light in this
example, many leptons do get produced from decays of the $W$ and
$b$ particles in $t\to bW$ decays.

The lepton multiplicity and lepton $p_T$ depend on the
$M_1$ and $M_2$ masses.  We illustrate this dependence by first
calculating lepton observables for our example model point 1, and
then doing the calculation for the same model but with $M_1$ and
$M_2$ redefined to be equal to $M_i=\alpha_iM_3/\alpha_s$, consistent
with gaugino mass unification, while $M_3$ remains the same.  In this
case, $M_1$ and $M_2$ are reset to $75\gev$ and $145\gev$ respectively,
and $M_3$ remains at 500~GeV.
In Table~\ref{lepton multiplicity} 
we list the total multiplicities of leptons in 1000 simulated
LHC events for the DAM example model, and the DAM example model with
$M_1$ and $M_2$ redefined. The lepton multiplicity is defined to be the
number of charged leptons of first and second generation with 
pseudo-rapidity $\eta < 3$ and 
transverse momentum $p_T>10\gev$
present in each supersymmetry event.  We have also required the
missing energy to be greater than $200\gev$ in these events to
reduce standard model background, and we have not counted leptons
that originate in a QCD jet (isolation requirement).

The lepton multiplicity
tends to higher values for the GUT normalized
gaugino spectrum rather than the
untampered DAM gaugino spectrum.  This is largely because more leptons
pass the $p_T>10\gev$ cut due to the large mass mass gap between the
mostly bino NLSP and the next higher mass chargino and neutralino.
If we put no cut on the $p_T$ of the lepton, the number of leptons
from the DAM model would be larger than the number of leptons generated
in the cascade decays of the GUT normalized gaugino version of the
spectrum.  (The number of leptons produced with arbitrarily low
$p_T$ values is listed in parenthesis in the table.)
This is indicative of the importance of looking carefully
at the $p_T$ spectrum of the leptons to see the imprint of 
different mass hierarchies
in the spectrum.

We demonstrate the softer lepton $p_T$ distributions of the 
DAM model in Fig.~\ref{pTlepton}.  We have 
simulated 1000 supersymmetric events at the
LHC and plotted the $p_T$ distribution of the leading lepton with
$\eta<3$ and $p_T>10\gev$.  The effect is present as
anticipated, and the magnitude of the effect is rather sizeable.  In the first
bin there is nearly a 50\% difference between the models.  We expect this
observable, along with other observables~\cite{Hinchliffe:1997iu}, 
such as kinematic
endpoint distributions,  to play a
key role in helping to distinguish DAM models from their competitors.
In this analysis we have been assuming that the signal with large missing
energy, large lepton multiplicity and large overall rate will
render the standard model background
not significant enough to diminish our conclusions, but of course a full
investigation of the background, and simulations of real detector
effects are necessary to make definitive statements about parameter
determinations in supersymmetric models. Nevertheless, we are encouraged
that distinctions between closely related models of supersymmetry 
can be made at hadron colliders.
\jfig{pTlepton}{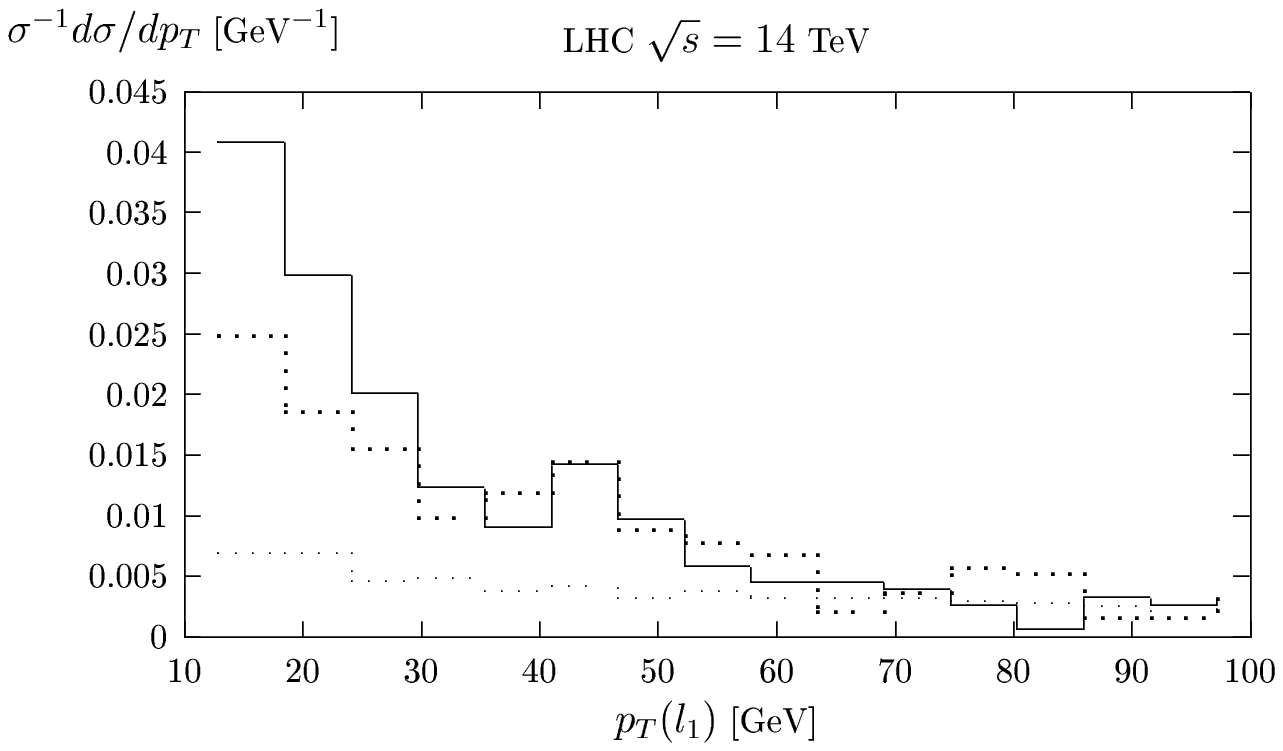}{The $p_T$ distribution of the leading lepton
in simulated events of supersymmetry production at the LHC. The solid
line is for DAM model point 1 described in the text. The dashed
line is for the same model except the electroweak gaugino masses are GUT
normalized with respect to the gluino ($M_i=\alpha_iM_3/\alpha_3$).
The fainter dotted line represents DAM model point 3.  All lines are normalized
to $1$ and not the total cross-section. (The total 
${\rm lepton}+X$ cross-section of DAM3 is a factor of 2.8 times that
of DAM1.)}

If $n\ge 6$ it is still possible to have Higgsino NLSP:
the change in the scalar masses of $H_u$ and $\tilde t_i$ with $\bar\lambda_H\sim 1$ can
alter the conditions for EWSB to allow a Higgsino NLSP.
For example, if we employ the same choices of parameters that we
used to generate DAM model point 2, except now we set
$\bar\lambda_H=1$, the resulting spectrum has a Higgsino NLSP.  This is model 
point 3 given
in the right two columns of Table~\ref{mtable2}.  The phenomenology of this
model with $n=6$ and Higgsino NLSP is dramatically different than the phenomenology of
point 1.  In contrast to DAM1, DAM3 has a high multiplicity
of leptons and high $p_T$ distribution of leptons.  
Table~\ref{lepton multiplicity}
lists the lepton multiplicities for model point 3, and the faint dotted curve
of Fig.~\ref{pTlepton} demonstrates the flat distribution of lepton $p_T$,
characteristic of a high $p_T$ spectrum of leptons.  These results
are readily understood by inspecting the mass hierarchies of point 3 compared
to point 1.  In point 3 the strongly interacting sparticles 
(squarks and gluinos)
will almost always cascade decay to a lepton.  The most effective path
is through $\tilde q\to \tilde W\to \tilde l$, where at least one
lepton results.  The mass hierarchies
of point 1 do not allow these high lepton multiplicities.  Therefore, 
close inspection of the lepton observables may provide a handle on
the parameter $\bar\lambda_H$ in addition to  measurements of the 
various sparticle masses.


\section{Signals in rare processes}
In DAM models the soft terms  could contain no extra flavour or CP violating terms
beyond the ones induced by the CKM matrix.
There are however two possible exceptions.
\begin{enumerate}
\item
The $\mu$ and $B$ terms could be complex: in this case they would
typically generate too large electron and neutron electric dipoles,
unless their phases are so small (less than about $0.01$)~\cite{dipoliSoft}
that do not significantly affect collider observables.

\item Extra Yukawa couplings not present in the SM
can affect the soft terms in a way that crucially depends
on how the soft terms are mediated.
In supergravity heavy particles affect the soft terms,
while in pure AM models soft terms are not affected by fields above the supersymmetry breaking scale.
Like in GMSB, in DAM models the soft terms are not affected by
interactions of fields heavier than the messenger mass $M_0$.
The effective theory at $M_0$  however might not be the MSSM.
For example, some of the `right-handed neutrinos' $N$
often introduced in order to generate the observed neutrino masses could be lighter than $M_0$.
If they have order one Yukawa couplings $\lambda_N \, NL\hu$ they
imprint lepton flavour violation in the soft mass terms of left-handed sleptons $\tilde{L}$
inducing significant rates for processes like $\mu\to e\gamma$.
Unlike in GMSB models, in DAM models
these effects are not suppressed by a (RG-enhanced) loop factor.
However for a right-handed neutrino mass 
$M_N\sim 10^{9\div 11}\GeV$, optimal for leptogenesis,
the Yukawa couplings $\lambda_N$
must be small, $\lambda_N\circa{<}0.005$, in order to get a left-handed
neutrino  mass smaller than $1\eV$.

\end{enumerate}
If none of these exceptions is realized,
in DAM models supersymmetric loop effects
only give new contributions to processes already present in the SM
($b\to s\gamma$, $g-2$ of $\mu$, $\epsilon_K$, $\Delta m_B$, $K\to\pi\bar{\nu}\nu$ decays)
but cannot give rise to new effects (like $\mu \to e \gamma$ decay, electric dipoles, contributions 
to $K,B_d,B_s,D$ physics with non-CKM and/or non-SM chiral structure).
Taking into account the accelerator bounds on sparticle masses,
few rare processes can receive interesting contributions:
\begin{itemize}

\item Supersymmetric corrections can significantly enhance 
${\rm BR}(B\to X_s \gamma)$~\cite{BBMR}
over its SM value.
For example in all the reference points studied in the previous
section the $b\to s\gamma$ effective operator
(with all fields and couplings renormalized at the relevant scale $Q\sim m_B$) is 
$$
{\cal H}_{\rm eff}=-[-0.29 ({\rm SM}) - 0.08 ({\rm charged~higgs}) \pm  0.07 ({\rm chargino})]
V_{tb}V_{ts}^*\frac{eg_2^2}{(4\pi)^2}\frac{m_b}{2M_W^2} [(\bar{s}_L \gamma_{\mu\nu} F^{\mu\nu} b_R)+\hbox{h.c.}]$$
Unless the chargino contribution compensates the charged higgs contribution
(its sign depends on the relative sign between $m$, $\mu$ and $B$),
${\rm BR}(B\to X_s \gamma)$ is two times larger than in the SM,
conflicting with experimental bounds.
Even assuming a destructive interference
(otherwise the sparticles must have unnaturally heavy masses)
a detectable supersymmetric correction to the $B\to X_s\gamma$ branching ratio
remains likely.
In these models the gluino/bottom contribution is computable, and turns out to be negligible.

\item 
Since EW gauginos are heavier than in mSUGRA or GM models,
a supersymmetric correction to the
anomalous magnetic moment of the $\mu$~\cite{gMuSUSY}, 
at a level detectable in forthcoming experiments~\cite{brook}
is rather unlikely (but not impossible).

\item The supersymmetric corrections to $K$ and $B$ mixing~\cite{BBMR}
can be larger than in mSUGRA and GMSB models, because
coloured sparticles can be lighter.
With a `reasonable' sparticle spectrum, $\Delta m_B$ can be enhanced by $(20\div 25)\%$
with respect to its SM value.
Such corrections are comparable to the present theoretical 
uncertainties on the relevant QCD matrix elements.
Larger corrections are present in small corners of the parameter space with light stops.
\end{itemize}


\section{NLSP decays and nucleosynthesis}\label{NLSP}

The lightest supersymmetric particle is the fermionic partner $\chi$ of the modulus $X$. Indeed 
by studying the effective action in eq.~(\ref{xaction})
one finds $m_\chi =\O(\alpha/4\pi)^2F_\phi$. 
Therefore, unless some coupling
in the messenger sector is strong, we expect $m_\chi$ to be smaller than a 
few GeV, so that $\chi$ is the LSP. The is a welcome fact: the LSP
of our model is automatically  neutral and unwanted
charged relics are avoided. 
On the other hand, the lightest sparticle in the SM sector, the NLSP, can
 be charged (a right-handed slepton)
as it decays into $\chi$. Now, the effective couplings governing this decay
 are suppressed by inverse powers of the messenger mass and by loop
factors.  Indeed $\chi$ plays a role similar to that of 
the Goldstino in gauge mediated models. In the range of allowed $M_0$, 
the NLSP lifetime is so long that it 
behaves as a stable particle in collider experiments. However, lifetimes in
excess of 1 sec, 
can dangerously affect the big-bang predictions of light element
abundances. In the rest of this section we will discuss the
constraints placed on $M_0$ by nucleosynthesis.
 
Let us first derive the couplings of $\chi$ to the
SM particles. 
For a chiral matter multiplet $Q$ the effective Lagrangian,
computing loop corrections with superfield techniques, is~\cite{PR}
\beq
\L_{\rm eff}=\int d^4\theta~Z_Q\left (\mu^2/\phi\phi^\dagger,
XX^\dagger/\phi\phi^\dagger\right ) QQ^\dagger
\label{lqq}
\eeq
leading to a coupling 
\begin{eqnsystem}{sys:Leff}
\L_{\chi q\tilde q}&=& \rho_q {F_\phi\over M_0} \chi q \tilde q^\dagger 
+{\rm h.c.}~~~~{\rm where,}\\
\rho_q&= &(\partial _{\ln \mu^2}+\partial_{\ln XX^\dagger})
\partial_{\ln XX^\dagger}\ln Z_Q.
\label{chiqq}
\end{eqnsystem}
The above expression is easily obtained by expanding $\ln Z_Q$ 
in powers of $\ln\phi$ and $ \ln X/M_0$ and 
by noting that the leading contribution to $\L_{\chi q\tilde q}$
comes from second order cross  terms $\propto\ln\phi \ln X/M_0$. In the
case of right-handed sleptons we have
\beq
\rho_{e_R} ={1\over 8\pi^2}{2n(n+33/5)\over 11}
\left (\alpha_1^2(M_0)-\alpha_1^2(M_Z)\right ),
\label{rhostau}
\eeq
where we have taken $\mu=m_Z$ in $\L_{\rm eff}$. Notice that the coupling
from eqs.~(\ref{chiqq}), (\ref{rhostau}) is qualitatively similar to the Goldstino
coupling for a gauge mediated model with $F_X/X=F_\phi$. However
in gauge mediation, unlike here, $\rho_q=m_{\tilde q}^2(M_0^2/F_X)^2$ by
current algebra. 

In the case of a higgsino NLSP the relevant term is the one generating 
$\mu$
\beq
\L_{\rm eff}=\int d^4\theta~\hu\hd {X^\dagger \over X}\tilde Z\left (
{\sqrt{XX^\dagger /\phi\phi^\dagger}}\right ).
\label{mueff}
\eeq
As discussed in section 2, the effective $\mu$ term is equal
to $(X^\dagger \tilde Z)|_{\bar\theta^2}/M_0$. By writing $X= M_0+\delta X$, 
it is easy to see
that, at the leading order in an expansion in $1/M_0$ and $\alpha$, 
eq.~(\ref{mueff}) leads to a superpotential coupling
\beq
\L_{\rm eff}=-\int d^2 \theta {\mu\over  M_0} {\hu\hd\, \delta X}.
\label{chihh}
\eeq
Notice that  the coupling of $\chi$ to the Higgs sector is stronger than
that to sfermions. It is 
 proportional to
the supersymmetric mass $\mu$ (1-loop) rather than to the mass splitting 
$B\mu$ (2-loop). This is consistent, since $\chi$ is not the Goldstino.
The most important consequence of  
eq.~(\ref{chihh})  is that it can mediate the decay $N_1\to h \chi$
whenever allowed by phase space. For a higgsino-like NLSP, we have 
$m_{N_1}\simeq \mu$ with 
$N_1\simeq \tilde H_\pm^0=(\tilde{H}_{\rm d}^0\pm \tilde{H}_{\rm u}^0)$, 
depending on the
sign of $\mu$. The width of a higgsino-like
NLSP is then
\beq
\Gamma_{N_1\to h\chi}={(\cos \alpha \mp \sin \alpha)^2\over 64 \pi}
{\mu^3\over M_0^2}\left(1-{\mu^2\over m_h^2}\right)^2
\eeq
corresponding to a lifetime shorter than a second over most of
parameter space already for $M_0\lsim 10^{15}$ GeV. We conclude that
nucleosynthesis does not place significant bounds on
a higgsino LSP whenever $\mu > m_h$, 
which is almost required by experimental bounds.
 
Let us consider now the bounds on a stau NLSP. 
The coupling to $\chi$ is smaller than for the higgsino NLSP (2-loop
versus 1-loop). The 
correspondingly longer $\tilde \tau$ lifetime is well 
approximated, as a function of $m_{\tilde \tau}$ and
$M_0$, by
\beq
\tau_{\tilde \tau}= \left ({M_0\over 10^{13} {\rm GeV}}\right )^2
\left ( {200 {\rm GeV}\over m_{\tilde \tau}}\right )^3\quad {\rm sec}.
\label{lifestau}
\eeq
This quantity is larger than 1 sec
over a significant fraction of parameter space, where the $\tilde{\tau}$ decay
can dangerously affect nucleosynthesis. 
 The most stringent bounds
come from decays processes involving hadronic showers. These showers
break up the ambient ${}^4 {\rm He}$ into D and ${}^3 {\rm He}$
and can lead to an overabundance of the two latter elements.  The showers
can also overproduce ${}^6 {\rm Li}$ and  ${}^7 {\rm Li}$ from
``hadrosynthesis'' of  ${}^3 {\rm He}$, T or  ${}^4 {\rm He}$. The decay
$\tilde \tau \to \chi \tau$ leads to hadronic showers as the $\tau$
further decays hadronically with a large branching ratio. Using the results
in ref.~\cite{rs,dehs}, it was concluded in ref.~\cite{ggr} 
that lifetimes larger than $10^{4}$ sec lead to unacceptable overproduction
of ${}^7{\rm Li}$. Ref.~\cite{ggr} shows a careful analysis,
including a computation of the relic NLSP density at nucleosynthesis,
for gauge mediated models with a stau NLSP. 
A similarly detailed analysis is beyond the aim of the present paper,
but we expect that the results of~\cite{ggr} can be carried over to
our case. This is because the bounds do not depend very
strongly on the $\tilde \tau$ relic density, which in our
model is not going to differ drastically from that in gauge mediation.
Therefore we conclude that overproduction of ${}^7 {\rm Li}$ gives the
bound $\tau_{\tilde \tau}< 10^4$ sec. By eq.~(\ref{lifestau}) this bound 
roughly translates into  $M_0<10^{14}$ GeV.

A stronger bound, forbidding decays between $10^2$ and $10^4$ sec can come 
from the 
deuterium abundance $X_D$ normalized to hydrogen. However there is,
at the moment a controversy in the measurement of $X_D$ from astrophysical
observation. Two values are quoted in the literature, a high one
$X_D=(1.9\pm 0.5)\times 10^{-4}$ from ref.~\cite{high} and a low
one $X_D=(3.39\pm0.25)\times 10^{-5}$ from ref.~\cite{low}. In ref.~\cite{ggr}
it was concluded that no further bounds are obtained when the high value 
of $X_D$ is assumed. On the other hand, the  low $X_D$ value
can give a stronger constraint $\tau_{\tilde \tau}< 10^2$ sec.

We conclude that nucleosynthesis places a significant bound on the
messenger mass when the NLSP is a stau. This bound on $M_0$ can range
between $10^{13}$ and $10^{15}$ GeV depending upon the model
parameters $m_{\tilde \tau}$ and $n$ and on the astrophysical input data.
We stress that while the bound is not negligible, there remains a large
allowed region    $ 10^{10}<M_0\lsim 10^{14} $ GeV, where nucleosynthesis 
is fine.

\section{Conclusions}

We have studied the phenomenology of models where the presence
of a light modulus $X$ induces a calculable correction to
anomaly mediated soft masses. This correction lifts the tachyonic sleptons
while preserving the flavor universality of anomaly mediation.
The resulting MSSM phenomenology is interesting and fairly distinguished
from both minimal supergravity and gauge mediation. The gaugino masses
are not unified, and the gluino is not much heavier than the other gauginos
(see fig.\fig{spettro}).
All sfermion masses start out negative at a scale between $10^{10}$ and
$10^{16}$ GeV but are driven positive at a lower scale by the RG contribution
of gaugino masses. Because of all these features the spectrum is 
a lot more compact than in minimal supergravity or gauge mediation
so that coloured sparticles can be produced and studied at TEVII.
Gauginos and squarks have comparable masses and are somewhat heavier than
higgsinos and sleptons.
The lightest superpartner is either a neutral higgsino
or a right-handed stau, but it is only an NLSP.
The LSP is the fermionic partner $\chi$ of the modulus $X$. 
The NLSP decay into $\chi$ takes place 
outside the detector. The rate of this decay does not conflict with 
the successful predictions of big bang nucleosynthesis over a significant
portion of parameter space.

The signals of deflected anomaly mediation at hadron colliders are easily 
distinguished from the conventional ones. In the case of a
charged slepton NLSP (e.g., DAM2) the signature is similar to GMSB with
two or more messengers:
two highly ionizing tracks in the detector. However,  the total production 
cross section as a function of the slepton mass is much 
bigger in DAM than in GMSB. For instance at Tevatron it could be a factor
20 bigger. This is because for a given slepton mass, gluinos
and squarks are about a factor of 2 lighter in DAM than in GMSB.
So in case stable charged tracks are discovered, one can easily tell
DAM from GMSB.

When the NLSP is higgsino (e.g., DAM1)
the competing scenarios have usually bino
LSP. Here the relevant observables are lepton multiplicity and 
$p_T$ distributions in supersymmetric events.  The signature of DAM depends
crucially on which is the bigger between the squark and wino mass
(each case can arise by proper parameter choices in DAM).
Since $m_{\tilde W}> m_{\tilde q}$ for DAM1, 
squark production leads to a cascade 
with fewer high $p_T$ leptons than in standard bino LSP scenarios. 
The softness of the leptons
is due to the small mass splitting among the charged and neutral
higgsinos produced in the cascade, while in mSUGRA and GMSB the LSP is
well split from the next higher mass neutralino and chargino.
Also, now the squarks often
decay directly to the lightest higgsino, without producing any lepton.

On the other hand for  $m_{\tilde W}< m_{\tilde q}$ (e.g., DAM3), more 
high $p_T$ leptons are produced
than usual. This is because squarks can decay via
$\tilde q_L\to \tilde W \to \tilde H^0_{1,2}, \tilde H^+$
and $\tilde q_L\to \tilde W\to \tilde l\to \tilde H^0$.  
Energetic leptons are then produced in $\tilde W$ decays
and/or the $\tilde l$ decays, while additional
softer leptons are produced when $\tilde H_2$ and
$\tilde H^+$ further decay to $\tilde H_1$. The lepton $p_T$ distribution
for the above cases is shown in 
Fig.~\ref{pTlepton} where it is compared to a 
standard bino LSP scenario, and lepton multiplicities are given in Table~3. 
Of course similar signatures are obtained in any
scenario where the higgsinos 
are somewhat lighter than winos and bino.  However, the unique mass hierarchy
of the charginos, neutralinos and sleptons 
in DAM, as illustrated by the spectrum of DAM3,
rarely occurs in GMSB or mSUGRA.
To further tell DAM from these other possibilities one can resort to
other observables. 
One additional consequence of the compact DAM spectrum is that
more than 2 tops in the gluino cascade are often forbidden by phase space,
whereas higher multiplicity of top quarks may exist
in final states of GMSB and mSUGRA.

We conclude that DAM provides an interesting alternative to conventional
soft term scenarios from both the theoretical and the
phenomenological point of view. This example also provides hope that
we may not have to wait for the LHC to discover superpartners:
TEVII may have enough luminosity and energy.

\paragraph{Acknowledgements}
We thank K.~Matchev, S.~Mrenna,
M.~Nojiri, and F.~Paige for useful and stimulating 
discussions. R.R. and J.D.W. wish to thank the ITP, Santa Barbara
for its support during part of this work (NSF Grant No.\ PHY94-07194).

\appendix

\begin{table}
$$\begin{array}{|c||c|ccccc|ccc|}\hline
i&\phantom{-}b_i&c_i^Q&c_i^U&c_i^D&c_i^L&c_i^E&\vphantom{X^{X^X}}
c_i^{\rm u}&c_i^{\rm d}&c_i^{\rm e}\\[0.5mm] \hline
1&{33\over5}&{1\over30}&{8\over15}&{2\over15}&\vphantom{X^{X^X}}
{3\over10}&{6\over5}&{13\over15}&{7\over15}&{9\over5}\\
2&1&{3\over2}&0&0&{3\over2}&0&3&3&3\vphantom{X^{X^X}}\\
3&-3&{8\over3}&{8\over3}&{8\over3}&0&0&{16\over3}&{16\over3}&0
\vphantom{X^{X^X}}\cr\hline
\end{array}$$
\caption{Values of the RG coefficients in the MSSM.\label{tab:bcude}}
\end{table}

\section{RG evolution of soft terms with non unified gaugino masses}
In this appendix we present semi-analytic solutions
for the one-loop RG evolution of the soft terms in presence of the large
Yukawa coupling of the top.
We give the soft terms at an arbitrary
energy scale $Q$,
starting from an arbitrary scale $M_0$ with arbitrary gaugino masses $M_{i0}$,
sfermion masses $m_{R0}^2$, $A$ terms $A_{g0}^{\rm f}$, $\mu$-term $\mu_0$,
and $B$-term $B_0$.
We do not assume unification of the gauge couplings.
Here $i=\{1,2,3\}$ runs over the three factors of the SM gauge group,
${\rm f}={\rm u,d,e}$, $g=1,2,3$ is a generation index and $R$ runs over all
the scalar sparticles ($Q_g,U_g,D_g,E_g,L_g,\hu,\hd$).
These formul\ae, obtained with superfield techniques~\cite{aglr,Zsusy},
are significantly simpler than equivalent ones already existing in the 
literature~\cite{RGEsemianOld}
because they never involve double integrals over the renormalization scale.
The running soft terms renormalized at an energy scale $Q$ are
\begin{eqnsystem}{sys:RGEsols}
M_i(Q)&=&M_{i0}/f_i\\
\mu(Q) &=& \mu_0\cdot y^{b^{\rm u}_1} E_h\\
B(Q) &=& B_{0} + 2 x_{i^1}^L M_{i0}  - b^{\rm u}_1  I'/b_t\\
A^{\rm f}_g(Q) &=& A^{\rm f}_{g0} + x_{1i}^{\rm f}(E) M_{i0}-b^{\rm f}_g I'(E)/b_t\\
m_{R}^2(Q) &=& m_{R0}^2+ x_{i^2}^R M_{i0}^2-b_R^t I -
{\textstyle\frac{3}{5} \frac{Y_R}{b_1}} I_Y
\end{eqnsystem}
where
$$t(Q)\equiv \frac{2}{(4\pi)^2}\ln\frac{M}{Q}\qquad
f_i(t(Q))\equiv \frac{\alpha_i(M_0)}{\alpha_i(Q)}
,\qquad
E^\alpha (t)\equiv \prod_i f_i^{c_1^\alpha/b_i}(t),\qquad
x_{i^n}^\alpha\equiv \frac{c_i^\alpha}{b_i}(1-f_i^{-n})$$
and $M$ is any scale.
All $b$-factors are simple numerical coefficients:
the $b_i$ are the coefficients of the one-loop $\beta$ functions, $\{b_1,b_2,b_3\}=\{33/5,1,-3\}$.
The $Y_R$ are the hypercharges of the various fields $R$,
normalized as $Y_{E}=+1$.
The $b_R^t$ coefficients vanish for all fields $R$
except the ones involved in the top Yukawa coupling:
$b_{\hu}^t=1/2$, $b_{\tilde{Q}_3}^t=1/6$ and $b^t_{\tilde{U}_3}=1/3$.
The factor $I_Y=(1-1/f_1)\Tr[Y_R m_{R0}^2]$ takes into account a small RG effect
induced by the ${\rm U}(1)_Y$ gauge coupling.
The $\lambda_t$ effects are contained in
\begin{eqnsystem}{sys:IeII}
I'&=&\rho[ A_{t0} +  M_{i0}X_{i}] \\
I&=&\rho (m_{Q_30}^2+m_{U_30}^2+m_{\hu 0}^2)+(1-\rho)A_{t0}^2
+\nonumber \\&&
-\big[(1-\rho)A_{t0}- M_{i0}\rho X_{i}\big]^2
+\rho M_{i0}^2 X_{i^2}+
\rho M_{i0}M_{j0} X_{ij}
\end{eqnsystem}
where $A_{t0}=A^{\rm u}_{30}$ is the top $A$-term at $M_0$ and
\begin{equation}
X(M_0,Q)\equiv \int_{t(M_0)}^{t(Q)} E^{\rm u}(t)dt,\qquad
X_{i^n}(M_0,Q)=\frac{\int E^{\rm u} x^{\rm u}_{i^n}~dt}{\int E~dt},\qquad
X_{ij}(M_0,Q)=\frac{\int E^{\rm u} x^{\rm u}_{i^1}x^{\rm u}_{i^1}~dt}{\int E~dt}
\end{equation}
All the integrals are done in the same range as the first one.
The `semi-analytic' functions $X_{i^n}$ are needed only for $n=1$ and $2$.
In practice one has to compute numerically few functions of two variables, $Q$ and $M_0$.
A more efficient computer implementation is obtained rewriting the $X_{\cdots}(M_0,Q)$ functions in terms
of $1+3+9$ functions with only one argument
$$F(M_0)-F(Q)=\int_{t(M_0)}^{t(Q)} E^{\rm u}(t)dt,\qquad
F_{i^n}(M_0)-F_{i^n}(Q)=\int {E^{\rm u}\over f_i^n}dt,\qquad
F_{ij}(M_0)-F_{ij}(Q)=\int {E^{\rm u}\over f_i f_j}dt .$$

\def\ijmp#1#2#3{{\it Int. Jour. Mod. Phys. }{\bf #1~}(19#2)~#3}
\def\pl#1#2#3{{\it Phys. Lett. }{\bf B#1~}(19#2)~#3}
\def\zp#1#2#3{{\it Z. Phys. }{\bf C#1~}(19#2)~#3}
\def\prl#1#2#3{{\it Phys. Rev. Lett. }{\bf #1~}(19#2)~#3}
\def\rmp#1#2#3{{\it Rev. Mod. Phys. }{\bf #1~}(19#2)~#3}
\def\prep#1#2#3{{\it Phys. Rep. }{\bf #1~}(19#2)~#3}
\def\pr#1#2#3{{\it Phys. Rev. }{\bf D#1~}(19#2)~#3}
\def\np#1#2#3{{\it Nucl. Phys. }{\bf B#1~}(19#2)~#3}
\def\mpl#1#2#3{{\it Mod. Phys. Lett. }{\bf #1~}(19#2)~#3}
\def\arnps#1#2#3{{\it Annu. Rev. Nucl. Part. Sci. }{\bf #1~}(19#2)~#3}
\def\sjnp#1#2#3{{\it Sov. J. Nucl. Phys. }{\bf #1~}(19#2)~#3}
\def\jetp#1#2#3{{\it JETP Lett. }{\bf #1~}(19#2)~#3}
\def\jhep#1#2{{\it JHEP } #2#1~(19#2)}
\def\app#1#2#3{{\it Acta Phys. Polon. }{\bf #1~}(19#2)~#3}
\def\rnc#1#2#3{{\it Riv. Nuovo Cim. }{\bf #1~}(19#2)~#3}
\def\ap#1#2#3{{\it Ann. Phys. }{\bf #1~}(19#2)~#3}
\def\ptp#1#2#3{{\it Prog. Theor. Phys. }{\bf #1~}(19#2)~#3}

\end{document}